# Size-dependent mass absorption cross-section of soot particles from various sources

Joel C. Corbin[1,*], Tyler J. Johnson[2], Fengshan Liu[1], Timothy A. Sipkens[1], Mark P. Johnson[3], Prem Lobo[1], Greg J. Smallwood[1]

[1]*Metrology Research Centre, National Research Council Canada, Ottawa, Ontario, Canada*
[2]*Atmose Ltd, Edmonton, Alberta, Canada*
[2]*Rolls Royce plc, Derby, UK*

**Abstract**
The mass absorption cross-section (MAC) of combustion-generated soot is used in pollution and emissions measurements to quantify the mass concentration of soot and in atmospheric modelling to predict the radiative effects of soot on climate. Previous work has suggested that the MAC of soot particles may change with their size, due to (1) internal scattering among monomers in the soot aggregate, (2) the correlation of soot primary-particle diameter with aggregate size, (3) quantum confinement effects, or (4) a size-dependent degree of soot graphitization. Here, we report a size-dependent MAC for ex-situ soot sampled from two commercially available diffusion-flame soot generators, one aviation turbine engine, and one diesel generator. We also incorporate literature data. We show that the MAC increases with aggregate size until a plateau is reached at single particle masses between 4 and 30 fg (approximately 300–650 nm soot mobility diameter). The smallest particles may have MACs 50% to 80% smaller than the largest, depending on the source, while the largest particles have MACs within the range reported by previous measurements on polydisperse samples. Moreover, we show that models of hypotheses (1), (2), and (3) do not describe our measurement results, leaving hypothesis (4) as the only remaining candidate.

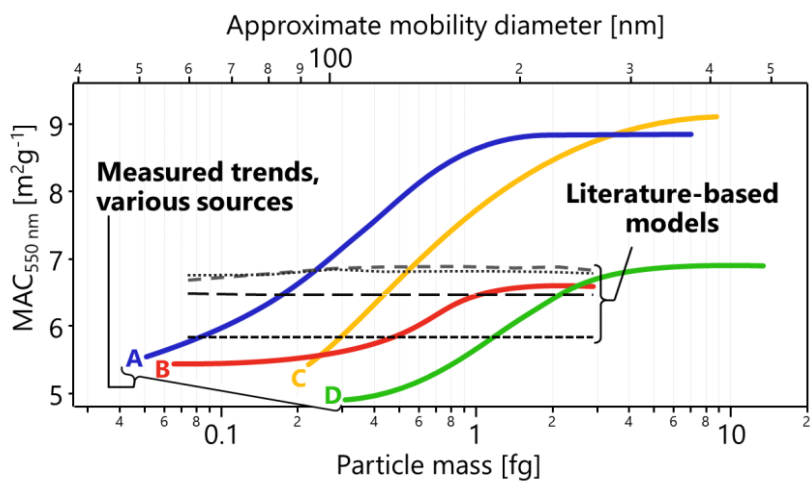

*Corresponding author: tel. 613-993-2176, Joel.Corbin@nrc-cnrc.gc.ca; Draft prepared for Carbon; Last modified 2022-02-14*

# 1. Introduction

The incomplete combustion of hydrocarbon fuels often results in the nucleation and growth of carbonaceous soot particles. These flame-generated nanoparticle aggregates of graphitized carbon spherules are highly light-absorbing; consequently, they play a major role in the radiative balance of the Earth [1] and in heat transfer in combustion [2]. Soot's light absorption is also commonly exploited for its identification and quantification [3] since soot particles generally dominate total light absorption by atmospheric aerosol particles [4] and often also in flames [5]. Conversely, soot's absorption properties are often used to predict its radiative effects from gravimetric or chemical measurements [1]. Therefore, accurate measurements and predictions of the impacts of the soot emitted from combustion require a thorough understanding of the factors influencing soot's light absorption.

Soot particles can also be called black carbon (BC) [1,6] or soot BC [7] to emphasize their strong broadband light absorption. In contexts such as regulatory measurements for automotive- [8] or aircraft-engine [9] emissions certification, soot particles are measured as non-volatile particulate matter (nvPM). Other forms of BC also exist, such as the carbonized fuel droplets or cenospheres often called char [7]. Soot particles are not purely composed of graphitized carbon; they contain approximately 2–10 % oxygen and a small amount of hydrogen in the form of refractory surface functional groups [10–13] and a smaller amount of metallic elements due to refractory metal compounds (ash) [14,15]. Some complexity is introduced by volatiles such as volatile organic compounds (VOCs) which often condense onto soot particles; these VOCs are not considered to be part of the soot particle [6] as they exist in a dynamic equilibrium with the gas phase.

Light absorption by soot particles at wavelength $\lambda$ is often quantified by the mass absorption cross-section ($MAC_\lambda$): the absorption coefficient of a population of particles normalized to their mass concentration. Unlike the refractive index (RI) or absorption function $E(m)$, which must be retrieved through model fits, MACs can be determined directly from measurements [16]. The MAC of soot aggregates may be influenced by their morphology [17–20] and molecular structure (degree of graphitization) [21,22]. The MAC may also be enhanced by coatings of non-absorbing materials such as VOCs [23–25]; this enhancement is complicated by the possibility of soot restructuring during VOC condensation.

Although comprehensive reviews of the typical MAC of soot have focussed on measurements with no size classification [16,26], size-dependent soot MACs have been measured in four previous studies. Dastanpour et al. [27] reported a size dependence of the MAC of soot particles produced by a large, inverted diffusion flame. They described this dependence using a power-law fit of MAC versus single-particle mass ($m_p$). Forestieri et al. [28] also observed a size-dependent MAC in a compiled laboratory data set of diffusion-flame soot. They found that a Rayleigh-Debye-Gans (RDG) model (which treats soot spherules as non-interacting primary particles) overpredicted the MAC of smaller aggregates ($x < 0.9$) but could adequately represent larger ones. They developed an empirical description of the MAC of smaller aggregates based on Mie theory (which treats aggregated soot spherules as volume-equivalent spheres) with an empirically fitted RI. Finally, Khalizov et al. [29] and Kholghy and DeRosa [30] also observed size-dependent MACs for diffusion-flame and flame-spray-pyrolysis soot, respectively. Two other studies have reported size-dependent MAC values for mature soot [31,32] but only measured large particles (with mass $m_p \gg 1$ fg) in the regime where negligible size dependence has been observed. Although most of these studies classified particles by mass rather than size, we use the term size here due to the fundamental importance of the size parameter $(x = \pi d_{pp}/\lambda$, where $d_{pp}$ is the monomer diameter$)$ in nanoparticle optics.

The RDG and Mie theories provide the simplest models of soot MAC. Mie models represent the entire soot aggregate as a volume-equivalent sphere and cannot describe the size-dependent MAC accurately [28]. RDG models are more accurate for fractal aggregates, but employ the assumption of negligible internal scattering among monomers [33]. The effect of neglecting monomer interactions on aggregate absorption can be considered by introducing an RDG correction factor $h$ derived from accurate numerical methods such as generalized multiparticle Mie (GMM) theory, the T-matrix method, or the discrete dipole approximation. The $h$ factor quantifies the influence of internal scattering on the soot MAC. By definition, $h$ is 1.0 for a single monomer. For a typical monomer diameter $d_{pp} = 15$ nm and visible wavelengths, $h$ first increases to a plateau at $h \approx 1.2$ for an aggregate composed of a moderate number of monomers ($10^1 < N_{pp} < 10^2$) in point contact, then decreases for larger aggregates $(N_{pp} > 10^2)$ [33]. We note that these values of $h$ represent specific values of size parameter $x$, RI, and the assumption of a diffusion-limited cluster aggregation (DLCA) soot morphology.

Optical models can also be used to quantify the influence of other details of soot structure on MAC, in addition to internal scattering. For example, Dastanpour et al. [27] observed a correlation between monomer size ($d_{pp}$) and aggregate size, and discussed its possible influence on MAC which can be readily modelled. Dastanpour et al. also measured a size-dependent soot maturity based on particle mobility diameter $d_m$, for $125 < d_m < 300$ nm particles using Raman spectroscopy [27,34] and attributed their observed size-dependent MAC to this trend. A quantitative model for the evolution of soot maturity in flames was subsequently described by Kelesidis and Pratsinis [35]. That model incorporated the quantum confinement effects identified by Wang and co-workers [36] for incipient soot particles in premixed flames $(4 < d_m < 23 \text{ nm})$, thereby parameterizing the refractive-index-dependent absorption function $E(m)$ in terms of $d_m$ via the electronic band gap $E_g$ [35]. This maturity model is independent of the internal scattering and monomer-aggregate size correlation hypotheses.

In addition to the above effects, the effects of monomer polydispersity and necking and overlap between monomers on aggregate absorption or MAC have also been investigated. The former effect (polydispersity of $d_{pp}$) has a negligible influence on the MAC of DLCA soot aggregates with monomers in point contact [37,38]. The latter effect (necking and overlap) also has only a small predicted influence on soot MAC, affecting $h$ by less than 10% in studies which modelled soot morphology phenomenologically, based on electron micrographs [19], or mechanistically, based on in-flame soot formation processes [39]. In flames, necking and overlap substantially affect $h$ only for freshly formed soot particles because surface growth (which is the driving force behind overlap and necking) become less important during the later stages of soot evolution [35]. After soot formation and before soot emission from flames, processes such as oxidation and annealing [40] may also affect necking, overlap, and soot maturity. Overall, in mature soot, polydispersity and surface growth are unlikely to have a major influence on the soot MAC. We therefore focus our following discussion on the effects of aggregate internal scattering, monomer-aggregate size correlation, and quantum confinement, which were hypothesized as potential causes of a size-dependent soot MAC.

In this work, we measured a range of size-dependent MACs for soot from two commercial soot generators, an aviation turbine engine, and a diesel generator. We used various instruments to quantify absorption coefficients. We used a centrifugal particle mass analyzer (CPMA) to classify particles by mass-to-charge ratio and an electrometer to quantify total charge and therefore total mass. To interpret the UDAC-CPMA data in terms of the average transmitted particle mass (or, equivalent size), we developed an iterative-average-charge approach and validated it using more complete calculations. We discuss our measurements in the context of similar literature measurements, as well as in the context of numerical models representing the relevant size-dependent phenomena discussed above.

## 2. Methods

### 2.1. Samples

Four soot sources were used in this work: an Argonaut miniature inverted soot generator (MISG), a Jing miniCAST model 5201c burner, a Gnome aviation turbine engine, and a diesel generator (Honda, Model EX4D, operated at 89% of maximum sustainable load) (Table 1). The MISG is an inverted diffusion flame and was operated with air and propane flows of 7.5 standard litres per minute (SLPM) and 0.0625 SLPM, respectively. Unpublished measurements by our laboratory, using the same technique described below, have shown that the MAC of soot produced by the MISG is insensitive to both the flow rate and fuel composition; the MISG MAC is also stable over the course of several months [41]. Therefore, only a single MISG setpoint was used in this study.

Table 1. Sources and sizes of the soot particles measured in this work. $d_{pp,100}$: mean spherule diameter for soot aggregates with $d_m \approx 100$ nm. CMD: count median diameter (equivalent to geometric mean diameter, GMD, for lognormal data). GSD: geometric standard deviation. MMD: mass median diameter of a lognormal distribution with the specified CMD and GSD neglecting density variation with size.

| Source | Fuel | $d_{pp,100}$ [nm] | CMD [nm] | GSD [-] | MMD [nm] |
|---|---|---|---|---|---|
| Gnome turbine | Jet A-1 | 20 [a] | 40 [b] | 1.8 [a] | 110 |
| Diesel generator | Diesel | - | 100 | 1.6 | 190 |
| Argonaut MISG | Propane | 18 [c] | 215 | 1.7 | 500 |
| miniCAST-S | Propane | - | 266 | 1.5 | 435 |
| miniCAST-D | Propane | 26 [d,*] | 250 [e] | 1.6 | 485 |
| miniCAST-A | Propane | - | 257 | 1.5 | 420 |

*Measurements from [a]Olfert et al. [46]; [b]Crayford et al. [69]; [c]Baldelli et al. [34]; [d] Durdina et al. [41]; [e]this study (Note that Durdina et al. [41] reported 150 nm). [*]Reported by Ref. [41] without reference to aggregate size.*

The miniCAST produces soot from a quenched diffusion flame. In contrast to the MISG, the MAC of soot produced by the miniCAST has been directly observed to vary with combustion-gas flow [41] as suggested by changes in related physical parameters [42–44]. We therefore operated the miniCAST at the three setpoints shown in Table 2. Setpoints D and S represent mature soot with a high degree of graphitization according to previous measurements in the literature [41,45]. Setpoint A mimics the conditions used to calibrate the Micro Soot Sensor plus (MSS) for aviation turbine engine non-volatile particulate matter emissions measurements.

Table 2. Combined summary of flow conditions for the Argonaut and miniCAST soot generators. miniCAST setpoints D, S, and A refer to the first setpoint characterized by Durdina et al. [41], the FL2 setpoint characterized by Saffaripour et al. [45], and the setpoint that has been used to calibrate the MSS for aviation turbine engine soot, respectively.

| Setpoint | Flow [SLPM] | | | |
|---|---|---|---|---|
| | Propane | Air | Premixed $N_2$ | Post-flame dilution air |
| Argonaut MISG | 0.075 | 7.5 | 0 | 0 |
| miniCAST-D | 0.071 | 1.91 | 0 | 9.8 |
| miniCAST-S | 0.055 | 1.6 | 0 | 20 |
| miniCAST-A | 0.060 | 1.6 | 0 | 20 |

Emissions from all the soot sources were sampled using heated lines (60 °C) and diluted with room-temperature air upstream of the catalytic stripper. In some cases, dilution factors were as low as 2 to ensure that signals were sufficiently high for measurements at the smallest concentrations, and to prevent water condensation in the sampling lines. Further details of the sampling configuration for the Gnome turbine engine at the Rolls-Royce facility in Derby, UK are available in Ref. [46]. The general laboratory setup for the measurements performed at the NRC has been described previously [47].

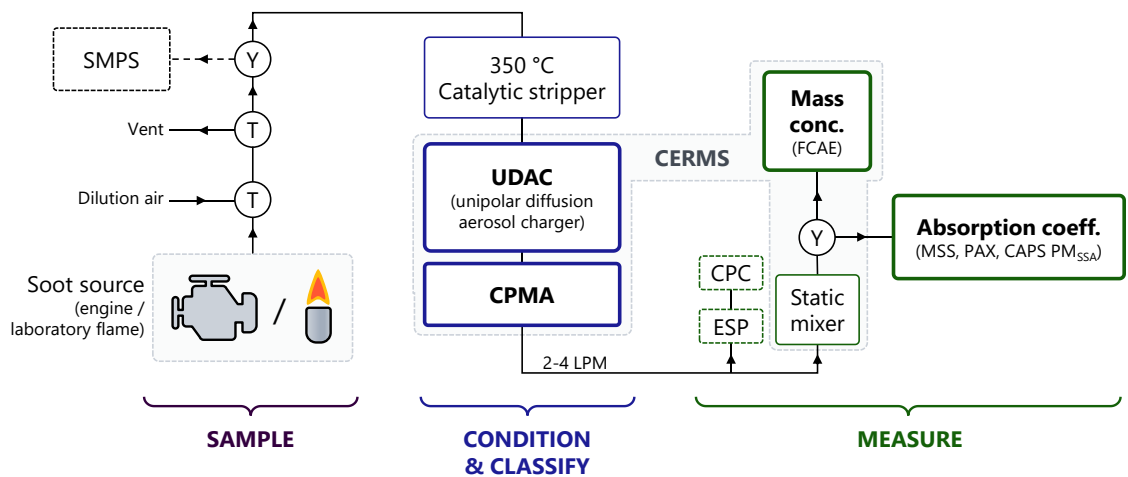

Figure 1. Experimental setup. A heated line (60 °C) was used between the engine or laboratory flame and the dilution stage. CPMA: Centrifugal particle mass analyzer. CPC: condensation particle counter. ESP: electrostatic precipitator. FCAE: faraday cup aerosol electrometer. MSS, PAX, and CAPS PM$_{SSA}$: instruments used for absorption coefficient measurements (see Table S1). SMPS: scanning mobility particle sizer; used for asynchronous measurement of the initial size distribution.

## 2.2. Soot characterization

Scanning mobility particle sizers (SMPSs; TSI Inc., USA) were used to characterize the size distributions of the non-volatile particles measured downstream of a 350 °C catalytic stripper (CS015, Catalytic Instruments GmbH, Germany). Below this temperature, the volatile mass fraction of MISG [47] and mature miniCAST soot (i.e. miniCAST soot without fuel dilution by premixing with N$_2$) [42,48] is negligible. Even at 600 °C, the volatile mass fraction of mature miniCAST soot is negligible [42].

The SMPS results are summarized in Table 1 in terms of the measured count median diameter (CMD) and geometric standard deviation (GSD). The mass median diameter (MMD) calculated from these properties [49] is also included. We also report the primary particle diameters in terms of $d_{pp,100}$ (the mean primary particle diameter of $d_m = 100$ nm aggregates) from literature for the Gnome [46] and Argonaut MISG [34] sources. For the miniCAST, no $d_{pp,100}$ measurements were available and we quote instead the $d_{pp}$ reported by Ref. [41]

for setpoint D, although this value was reported without reference to aggregate size. We expect that the miniCAST $d_{pp}$ is biased high (i.e., represents the mean primary particle diameter of aggregates with $d_m >$ 100 nm, instead of $d_m =$ 100 nm as assumed) considering the normal operator bias towards larger particles and the shift of the aggregate size distribution towards larger sizes.

## 2.3. Mass absorption cross-section (MAC) measurements

The MAC of a population of particles [units: $m^2 g^{-1}$] is evaluated as

$$\text{MAC}_\lambda = \frac{B_{\text{abn},\lambda}}{M_{\text{PM}}} \qquad 1$$

where $\text{MAC}_\lambda$ is the MAC measured at wavelength $\lambda$, $B_{\text{abn},\lambda}$ is the absorption coefficient at wavelength $\lambda$, and $M_{\text{PM}}$ is the particulate matter (PM) mass concentration. The quantity $M_{\text{PM}}$ represents the soot mass concentration, since we measured non-volatile PM from the combustion of hydrocarbon fuels with negligible metal and sulphur content.

Equation 1 is typically used to measure the MAC of all particles produced by a given source. For a source producing particles with a *size-dependent* MAC, Equation 1 may be formulated as an integral to give the *size-integrated* MAC:

$$\overline{\text{MAC}_\lambda} = \frac{1}{M_{\text{PM}}} \sum_i \frac{\Delta B_{\text{abn},\lambda,i}}{\Delta M_{\text{PM},i}} \qquad 2$$

$\overline{\text{MAC}_\lambda}$ is the quantity measured by most previous studies, including those reviewed by Refs. [16,26]. In Equation 2, Δ indicates a subset of the particles, $M_{\text{PM}} = \sum_j \Delta M_{\text{PM},i}$ and $B_{\text{abn},\lambda} = \sum_j \Delta B_{\text{abn},\lambda,i}$, for example, as produced by the UDAC-CPMA system described below. We note that the UDAC-CPMA system physically classifies particles by mass, not size, but we use the term size-integrated to avoid confusion as discussed in Section 3.

Since we used various instruments to measure $B_{\text{abn},\lambda}$ at different $\lambda$, we followed

Refs. [1,16,26] in converting MAC at measured the wavelengths to an equivalent MAC at 550 nm ($MAC_{550nm}$) as

$$MAC_{550nm} = MAC_\lambda \left(\frac{550\text{nm}}{\lambda}\right)^{-AAE} \quad\quad 3$$

and assuming an absorption Ångström exponent (AAE) of 1. We used the same conversion for literature data. Our procedure for MAC measurements is summarized in the following subsections and in Table S1.

## 2.4. CERMS measurements of $M_{PM}$

We measured $M_{PM}$ using the CPMA-Electrometer Reference Mass Standard (CERMS) [47,50–52] shown in Figure 1. In the CERMS, sample particles are highly charged by a unipolar diffusion aerosol charger (UDAC, Cambustion, UK) before being transmitted to a centrifugal particle mass analyzer (CPMA, Cambustion, UK) and then a faraday cup aerosol electrometer (FCAE, TSI 3068B, TSI Inc., USA). The fraction of uncharged particles exiting the UDAC was verified to be negligible in a dedicated experiment using an electrostatic precipitator (ESP, Cambustion, UK) and condensation particle counter (CPC, TSI 3776, TSI Inc., USA) in series.

During $M_{PM}$ measurement, charged particles flow through the CPMA, which transmits particles of a known mass-to-charge ratio ($m/q$) to an electrometer and a challenge instrument. Particles with $m/q$ greater or smaller than the CPMA setpoint are removed by electrical or centrifugal forces. The expression $m/q$ refers to all particles with charge $q$ and mass $m_q = q \cdot m_1$, where $m_1$ is the mass of a singly charged particle. The actual distribution of particles transmitted by the CPMA will depend on the upstream mass and charge distribution (cf. Section 2.6. and Supplementary Information). Knowledge of this distribution is not necessary for mass quantification in the CERMS, because the electrometer measures total charge and because $m/q$ is fixed. Therefore, for an electrical current ($I$) produced by CPMA-classified particles at a volumetric flow rate $Q$, $M_\text{PM}$ can be calculated as [50]:

$$M_\text{PM} = \frac{m_1 I}{Qe} \qquad 4$$

where $m_1$ is the mass of a singly-charged particle and $e$ is the elementary charge. This $M_\text{PM}$ may be used to calibrate instruments reporting mass concentration [47,51], or can be used to calculate MAC if the challenge instrument measures $B_{\text{abn},\lambda}$. Based on our previous work [47], we rejected data where the electrometer measured less than 250 fA as inaccurate.

## 2.5. Measurements of $B_{\text{abn},\lambda}$

We calibrated and used three different instruments to measure $B_{\text{abn},\lambda}$ (Table S1). We used a CAPS PM$_\text{SSA}$ monitor (Aerodyne Research Inc., USA), which measures the lifetime of photons in an optical cell to derive the aerosol extinction coefficient $B_{\text{ext},\lambda}$ while an integrating nephelometer simultaneously measures the aerosol light scattering coefficient $B_{\text{sca},\lambda}$ from the same cell. The difference, extinction minus scattering, is then the absorption coefficient. The $B_{\text{ext},\lambda}$ measurement is calibration-free with an estimated accuracy of 5 % [53]. The $B_{\text{sca},\lambda}$ measurement is calibrated by reference to the $B_{\text{ext},\lambda}$ measurement using non-absorbing particles with diameters at least a factor of 3 smaller than the measurement wavelength [54]. Truncation uncertainty in the CAPS PM$_\text{SSA}$ nephelometer is negligible for the sizes measured here [54] and the overall uncertainty in $B_{\text{abn},\lambda}$ is estimated as approximately 10 % [54,55]. In this study, we size-classified non-absorbing (NH$_4$)$_2$SO$_4$ particles with an aerodynamic diameter of 290 nm (corresponding to an approximate mobility and spherical-equivalent diameter of 200 nm) using an Aerodynamic Aerosol Classifier (AAC, Cambustion, UK) and a measurement wavelength of 660 nm. At no point does this CAPS PM$_\text{SSA}$ calibration rely on a MAC.

The other two instruments used in this study were both photoacoustic spectrometers which periodically laser-heat particles in an acoustic resonator to generate pressure waves that are proportional to $B_{abn,\lambda}$ in magnitude. A microphone measures these pressure waves. The photoacoustic extinctiometer (PAX, Droplet Measurement Technologies Inc., USA) uses an 870 nm diode laser and reports $B_{abn,\lambda}$ as well as absorption-equivalent BC mass concentration ($M_{eBC}$) by applying Equation 1 with an assumed $MAC_{870nm}$ of 4.74 m$^2$ g$^{-1}$ (derived from $MAC_{550nm}$ of 7.5 m$^2$ g$^{-1}$ and AAE of 1.0).

The PAX is calibrated in two steps using two additional detectors, a laser power meter and a light-scattering detector. The laser power meter allows $B_{ext,\lambda}$ to be calculated at extremely high values ($> 10^4$/Mm) using the Beer-Lambert law. For non-absorbing particles, $B_{ext,\lambda} = B_{sca,\lambda}$. Thus, the first calibration step uses a high concentration of ammonium sulphate particles to define $B_{sca,\lambda}$ using the Beer-Lambert law. In the second calibration step, the ammonium sulphate is replaced with a high concentration of absorbing particles (we used Aquadag colloidal graphite) for which $B_{ext,\lambda} = B_{sca,\lambda} + B_{abn,\lambda}$. Since $B_{ext,\lambda}$ is known from the Beer-Lambert law and $B_{sca,\lambda}$ is known from the first step, a reference value of $B_{abn,\lambda}$ may be calculated and used to calibrate the PAX microphone. At no point does this PAX calibration rely on a MAC. Overall, this two-step PAX calibration requires the assumption that the instrument behaves linearly from $> 10^4$/Mm down to the normal measured range ($\approx 10$/Mm); an assumption that was recently verified experimentally [47]. By reference to an independent CERMS calibration, the PAX accuracy after calibration is estimated to be better than 20%, depending mainly on the rate of change of humidity in the sample aerosol [47].

The MSS (AVL, Austria) is a photoacoustic spectrometer similar to the PAX, but uses an 808 nm laser and does not incorporate additional detectors for internal calibration. Instead, it is calibrated by taking filter samples in parallel to the instrument. Either the gravimetric mass or the elemental carbon mass (EC, determined by thermal-optical analysis) on the filters is then used as a calibration reference, depending on whether the calibration is intended for use in the aviation sector for regulatory measurements of nvPM (EC) or not (gravimetric). Thus, in the aviation sector the MSS reported mass represents $M_{eBC}$ calibrated to the mass concentration of EC and also represents the assumption that the MAC of the calibration source and measured sample are equal. The difference between EC and PM mass concentrations is discussed in detail in [47]. The assumption that the MAC did not change is typically reasonably accurate for mature soot [16,26] but not for the present work. Here, we transformed the $M_{eBC}$ reported by the MSS into $B_{abn,808nm}$ by using the MAC of the Argonaut MISG as a reference as follows. First, the size-resolved $MAC_{870nm}$ of MISG soot was measured at NRC using the PAX and converted to $MAC_{808nm}$ using an AAE of 1. Then, the Argonaut MISG $M_{PM}$ was measured using the CERMS, and finally Equation 1 was applied to calculate the size-resolved $B_{abn,808nm}$ measured by the MSS. Our assumption that the Argonaut MISG MAC was constant is justified by literature reviews of soot in general [16,26] and by Argonaut-MISG-specific measurements performed in our laboratory [56]. This MSS calibration was performed on a size-resolved basis to avoid any assumptions on its size dependence. Since the resulting MSS data was not different from corresponding PAX and CAPS data, we consider the MSS calibration to have resulted in negligible additional uncertainty.

*2.6. Average single-particle mass of UDAC-CPMA classified particles*

In our experimental setup (Figure 1), particles are charged by the UDAC, configured to a specific ion-concentration-time product $n_i t$, before being classified by the CPMA. Under our measurement conditions ($n_i t > 10^{13}$ ion · s · m$^{-3}$ and room temperature and pressure of 298 K and 101.325 kPa), the UDAC imparts several charges ($q \gg 5$) to most particles. The CPMA transmits these particles at a specific mass-to-charge ratio, $m/q$, for all $q$ found in the upstream particle distribution. For example, $q = \{1, 2, 3, ...\}$ charges result in classified particle masses of $m_q = \{m_1, 2m_1, 3m_1, ...\}$, each of which may correspond to a different mobility diameter. More precisely, for each charge, the CPMA transmits a narrow lognormal distribution of particles with average mass $m_q$, where each distribution is given by the CPMA transfer function [57]. The net CPMA output may be viewed as a superposition of these lognormal transfer functions. Computing the CPMA transfer function using the finite-difference method of Ref. [57], we found that these lognormal transfer functions overlap with a relative intensity dependent on the aerosol charge and mass distributions (see Figure S1) for our conditions. Consequently, the sum of these overlapping lognormal transfer functions is itself approximately lognormal, centred about an average transmitted particle mass $m_p$, which depends on the average charge per particle $\bar{q}$ and the CPMA setpoint $m/q$:

$$m_p = (m/q) \times \bar{q}. \qquad 5$$

Based on Equation 5, we developed a novel and simple approach to correct for multiple charging in the UDAC-CPMA: the iterative-average-charge (IAC) algorithm, illustrated in Figure 2a. The algorithm is initialized by assuming $\bar{q} = 1$ in Equation 5 to provide an initial estimate of $m_p$. Then, the estimated $m_p$ is used to compute $d_m$ according to:

$$\rho_{\text{eff}} = \frac{m_p}{\pi d_m^3/6} = \rho_{\text{eff},100} \left(\frac{d_m}{100 \text{ nm}}\right)^{\varepsilon-3} \qquad 6$$

Here, the first equivalency defines the effective density and the second equivalency is a power-law fit (sometimes referred to as a fractal scaling law) which describes the effective density function of soot aggregates in terms of the effective density of particles with $d_m = 100$ nm, $\rho_{\text{eff},100}$, and the mass-mobility exponent $\varepsilon$. We used the power-law fit with $\rho_{\text{eff},100} = 510 \pm 8$ kg m$^{-3}$ and $\varepsilon = 2.48$ following the "universal" parameterization of Olfert and Rogak [58]. That parameterization represents soot from all of our sources except the diesel generator (Table 1) as well as various diesel engines. Equation 6 can also be formulated as a mass-mobility relationship:

$$d_m = (100 \text{ nm}) \left(\frac{m_p}{0.2670 \text{ fg}}\right)^{1/2.48} \qquad 7$$

where the constants are taken from the universal soot parameterization [58] as 0.2670 fg for the mass of a 100 nm mobility diameter soot particle with effective density 510 ± 8 kg m$^{-3}$ [58].

After $d_m$ estimation, the effective charging diameter $d_C$ is approximated as $d_C \approx d_m$, based on Ref. [59]. From $d_C$ and the UDAC $n_i t$ setpoint, the Fuchs charging model [60] can be used to predict a new value of $\bar{q}$, which, in turn, is used to update the previous estimate of $m_p$. This calculation is iterated until $m_p$ converges (Figure 2a).

The iterative-average-charge approach implicitly assumes that the average value of $m_p$ can be inferred from the average value of $q$. We evaluated this assumption by performing a more complete calculation, where we represented the mass distribution of the input particles (based on SMPS measurements; Table 1) and the charge distribution imparted by the UDAC onto particles at the CPMA setpoint ($m_q$). By combining these with the CPMA transfer function, we predicted the post-CPMA mass distribution (Figure 2b) from which the average single-particle mass can be directly computed (Equation S3). The resulting average-single-particle mass results were very similar (Figure S2; data points overlap for the two algorithms). This validates the simpler iterative-average-charge algorithm for our conditions. We also evaluated uncertainty in the input parameters to this model and associated assumptions, as discussed briefly in Section 2.7. and thoroughly in Section S1.3 of the supplementary information.

Finally, we also extended our interpretation of the post-CPMA mass distributions by implementing a system-of-equations approach to interpret the mass-and-charge distributions. In this approach, we allow for the possibility that particles with a range of MACs are transmitted by the UDAC-CPMA at a single setpoint (due to their range of charge states). The results of the system-of-equations approach were similar to those of the other two approaches (Figure S1 and Section S1.2), which is consistent with our earlier observation that the post-CPMA mass distribution was approximately lognormal and spanned a narrow range. For simplicity, we focus below on the results of the mass-and-charge distribution results. Further details of the charge correction are given in Section S1.2 and Figure S1, Figure S2, Figure S3, and Figure S4.

## *2.7. Uncertainties*

Uncertainties in our analysis can be separated into experimental uncertainties in the determination of MAC (corresponding to Figure 1) and uncertainties in our estimation of $m_p$ (corresponding to Figure 2). Experimental uncertainties in MAC were propagated through Equations 1 and 4, starting from the standard error of repeated measurements. These uncertainties are illustrated by representative error bars in our figures (uncertainties for each data point are also reported in the online supplementary information). Uncertainties in Equation 3 were assessed as negligible based on the consistency of results obtained for different sources; they also do not affect any of our conclusions.

Uncertainties in calibration constants were smaller than propagated measurement uncertainties, but may not account for instrument-specific biases, as discussed in Section 2.5. Our interpretation of $M_{\text{PM}}$ as accurately representing only non-volatile soot mass in Equation 4 was discussed in Section 2.2.

Uncertainties in $m_p$ estimation were assessed through perturbation of the input parameters (thick-lined boxes) in Figure 2. These uncertainties were dominated by uncertainty in the mass-mobility exponent (in particular for values significantly different from $d_m$ = 100 nm, which is the reference $d_m$ for the expression) of Equation 7 and are on the order of 20-30% depending on the absolute value of $m_p$ (Table S3) and on whether uncertainty in soot particle morphology is substantial. The details are discussed further in the supplementary information.

We emphasize that uncertainties in MAC and in $m_p$ are independent. The two quantities are used as the ordinate and abscissa axes below, but are not combined numerically. Any bias in $m_p$ (or soot mobility diameter $d_m$) will only change the apparent size of the observed trends in MACs.

*2.8. MAC modelling*

We modelled the MAC of soot using Mie, RDG, and GMM approaches to reproduce the literature hypotheses and models discussed in the introduction. The models and their inputs are summarized in Table 3. The models are simplified whenever possible due to the computational burden of the most precise (i.e., GMM) calculations.

Table 3. MAC model descriptions. Variables such as soot density which were common between models are omitted. DLCA: diffusion-limited cluster aggregation. All RIs were chosen for consistency with Ref. [35], since Model RDG-3 is based on the RI parameterization of that work. RDG refers to the Rayleigh-Debye-Gans approximation for aggregates and GMM to the General Mie model for aggregates.

| Model | Hypothesis tested | Soot morphology | $d_{pp}$ | RI |
|---|---|---|---|---|
| Mie | None (literature context only) | Volume-equivalent sphere | Fixed | $1.66 + 0.76i$ |
| RDG-0 | Null hypothesis for GMM-1 and RDG-3 | Non-interacting monomers | Fixed | $1.66 + 0.76i$ |
| GMM-1 | Internal light scattering | DLCA aggregates | Fixed | $1.66 + 0.76i$ |
| GMM-2 | Monomer–aggregate size correlation | DLCA aggregates | $f(d_m)$[a] | $1.66 + 0.76i$ |
| RDG-3 | Quantum confinement effects | Non-interacting monomers | Fixed | $= g(d_m)$[b] |

[a]Parameterized as a function of $d_m$ using Equation 11; resulting in $d_{pp}$ of 16 to 27 nm for the $d_m$ range 60 to 262 nm (cf. Figure S6). [b]Parameterized as a function of $d_m$ following Ref. [35] resulting in RI of $1.66 + 0.75i$ to $1.66 + 0.76i$, as discussed in the text.

Our initial model represented the aviation turbine engine soot described in Table 1 with RI of $1.66 + 0.76i$ [19], soot density $\rho = 1800$ kg m⁻³ [61], and $d_{pp,100} = 20$ nm [46]. This RI leads to calculated MACs that are close to the lower limit of the range of measured MACs as reviewed in Ref. [16]. It was chosen here for consistency with the model reported in Ref. [35] (see below). We converted the mass of soot particles to $d_m$ using the soot effective-density relationship described in Section 2.6.

For Mie calculations, we used a volume-equivalent diameter based on $m_p$ and the material density of soot (1800 kg m⁻³), defined as $[6m_p/\rho\pi]^{1/3}$ (*Mie model; used for literature context only*).

For RDG calculations, we used $d_{pp}$ as the particle size (*RDG-0 model*). According to the modified RDG approximation,

$$\mathrm{MAC}_{\lambda,\mathrm{RDG}} = h \cdot \frac{6\pi E(m)}{\lambda \rho} \qquad 8$$

where the absorption function $E(\mathrm{RI})$ is written as $E(m)$ according to convention, and is defined as

$$E(m) = \mathrm{Imag}[(\mathrm{RI}^2 - 1)/(\mathrm{RI}^2 + 1)] \qquad 9$$

In Equation 8, $\rho$ is the particle density, $\lambda$ the light wavelength, and $h = \mathrm{MAC}_{\lambda,\mathrm{accurate}}/\mathrm{MAC}_{\lambda,\mathrm{RDG}}$ the correction factor which scales the RDG MAC to the value predicted by accurate calculations. While Equations 8 and 9 fully describe our RDG-0 model, we note that these equations can also be related to the optical band-gap energy $E_g$ via parameterizations such as [35]:

$$\mathrm{RI} = (1.77 - 0.43 E_g) - (1.07 - 1.23 E_g)i \qquad 10$$

This parameterization is defined for an $E_g$ of 0.25 eV and 0.6 eV for mature and young soot, respectively [35], and results in a mature soot RI of $1.66 + 0.76i$.

Other models were implemented as follows. To account for the effects of internal scattering within a soot aggregate on MAC (which leads to $h > 1$ for aggregates of relatively small to moderate sizes), we used the GMM method for aggregates formed by point-contact monomers with $d_{pp} = 20$ nm as described previously [62] (*GMM-1 model*).

We also conducted GMM calculations for aggregates formed by DLCA with monomers whose diameter increases with the aggregate size [63], following the monomer-aggregate size correlation (*GMM-2 model*) reported in Equation 9 of Olfert and Rogak [58]:

$$d_{pp} = (d_{pp,100}) \left(\frac{d_m}{100\ \mathrm{nm}}\right)^{0.35} \qquad 11$$

where $d_{pp,100}$ was taken as 20 nm (Table 1). Consistent with Ref. [33], our DLCA aggregates have a fractal dimension of 1.78 and prefactor of 1.35.

To evaluate the quantum confinement and soot maturity effects on soot RI and ρ we used an RDG model (*RDG-3 model*) together with the optical band gap ($E_g$), RI, and ρ parameterizations given in Ref. [35]. Here, we used an RDG model because our GMM results had indicated a negligible influence of aggregation over the range of modelled aggregate sizes. The RI used in this model were calculated using Equation 10 with $E_g$ specified by Eq. 3 of Ref. [35].

We emphasize that $E_g$ also represents soot maturity, independently from the RDG-3 model. We have modelled our results in terms of MAC rather than $E_g$ because our measurements can be used to calculate MAC with no assumptions (Equation 1). In contrast, the calculation of $E_g$ from our measurements requires an assumed or parameterized soot density (Figure S5)$E_g$ via RI. Consequently, we focus the following discussion on MAC values and mention $E_g$ only in the context of soot maturity.

## 3. Results and Discussion

In the following discussion of MAC, we discuss only particles classified by single-particle mass. For discussion only, we converted single-particle mass $m_p$ to a soot mobility diameter $d_m$ using Equation 6. This conversion allows an illustrative upper axis to be added to our figures, but does not influence any of our reported trends, which are all plotted against $m_p$ except the illustrative Figure 4. The conversion allows us to simplify our discussion to avoid confusion between single-particle mass and quantities related to the integrated mass concentration (as in "mass absorption cross section") as well as to emphasize the dimension of size, which is important in optics as mentioned above.

## 3.1. Measured MAC from flames and engines

Figure 3 shows the dependence of the size-dependent MAC on single-particle mass ($m_p$) of soot produced by the Gnome aircraft turbine engine, the MISG (3 different absorption instruments), the miniCAST (3 different flame conditions), and the diesel generator. The diesel generator output was found to vary substantially from day to day, so repeated measurements on two days (day 1 and day 2) are labelled as such in the figure.

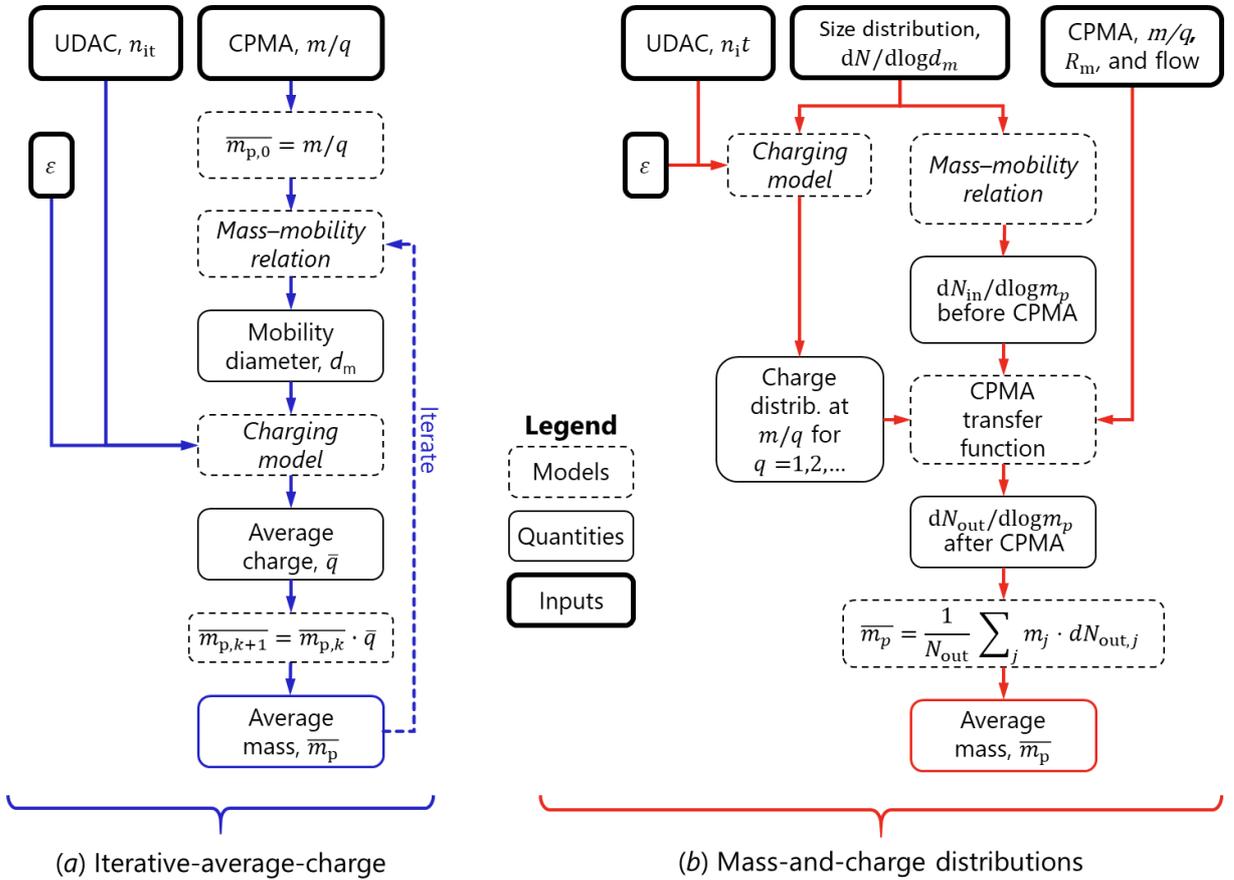

Figure 2. Schematic of the (a) iterative-average charge and (b) mass-and-charge distributions approaches to multiple-charge correction in the UDAC-CPMA. In both approaches, the mass-mobility relationship may be based on source-specific measurements or on a general parameterization [58]. The mobility diameter $d_m$ is used as input to the charging model as an approximation to the charging diameter. $\kappa$ is the dielectric constant of the particles. Note that the Fuchs model and the CPMA transfer function both depend on temperature and pressure.

All of the data suggest that a plateau of MAC exists at ~ 4 fg < $m_p$ < 30 fg (about 300 < $d_m$ < 650 nm, according to Equation 7) and that the MAC decreases below 4 fg. MAC is also likely to decrease above 30 fg based on Mie [26] or fractal aggregate [33] models, but this is outside the scope of our discussion where we focus on the more commonly observed soot sizes.

The data in Figure 3 span all $m_p$ for which the electrometer current was above the 250 fA limit of detection for the CERMS. Therefore, the range of reported $m_p$ represents the MAC of all particles present at substantial integrated mass concentrations $M_{PM}$. This is illustrated by the shaded curves in the lower part of Figure 3, which approximately represent the mass distributions of the Gnome aircraft turbine engine (grey) and the miniCAST-A (blue) samples. These approximate mass distributions were calculated by transforming the lognormal mobility size distribution parameters in Table 1 to mass distributions using the universal mass-mobility relationship of Olfert and Rogak [58].

For most sources, the functional form of the data in Figure 3 is not well-constrained by our measurements. The aircraft turbine engine data are an exception, and suggest a sigmoidal function (such as Equation S6). Extrapolating this sigmoidal function to all sources suggests a variation in the inflection point and slope of that function. This point is discussed further below.

### 3.2. Impact of size-dependent MAC on overall MAC of lognormal distributions

Soot sources generally produce polydisperse, lognormally distributed mobility-size or mass distributions. The dependence of the corresponding mass-integrated MAC, $\overline{MAC_\lambda}$ on the median of the particle size distribution will be similar to, but a smoother function of $m_p$ than that of the mass-classified soot shown in Figure 3, since the CPMA output represents a narrower range of sizes than a polydisperse distribution. There is no dependence of MAC on the median size for $m_p > 4$ fg (approximately $d_m > 300$ nm for soot according to Equation 7), where the size dependence of the MAC reaches a plateau. For most of our samples, the mass median diameters (MMDs) of $d_m = 400$ nm to 500 nm lie within this plateau regime. For instance, for a MMD of 400 nm and GSD of 1.6, 84% of the integrated lognormal mass is found in particles larger than MMD ÷ GSD = 250 nm [49], i.e., in the upper plateau region of Figure 3. In contrast, the aviation turbine engine and diesel generator have smaller MMDs of 110 nm and 190 nm, respectively. These MMDs fall in the region where the size dependence of MAC is the strongest. Therefore, measurements of such engine emissions, which report $M_{PM}$ by assuming a constant MAC (Equation 1), may introduce a size-dependent bias when assuming a size-independent MAC. In specific cases, this bias may be influenced by size-dependent processes such as sampling-line penetration losses.

Our assumed GMDs of 40 nm and 100 nm for aviation turbine engine and diesel generator soot are both above the midpoint of the range of sizes expected for such engines, particularly for modern engines. A recent review of aviation turbine soot [64] reported a range of GMDs from 15 nm to 60 nm, with smaller particles measured at lower thrust settings (<40% of maximum). GMDs of soot particles emitted from automotive gasoline-direct-injection (GDI) or port-fuel-injection (PFI) engines are similarly small; vehicles with model years after 2008 have GMDs from 10 to 70 nm (mean of type approval cycle data) [65]. (For earlier model years, the reported GMDs were progressively larger, up to a maximum of 100 nm.) Approximately half of the soot from such engines would be expected to fall close to the midpoint of the various size-dependent MACs shown in Figure 3.

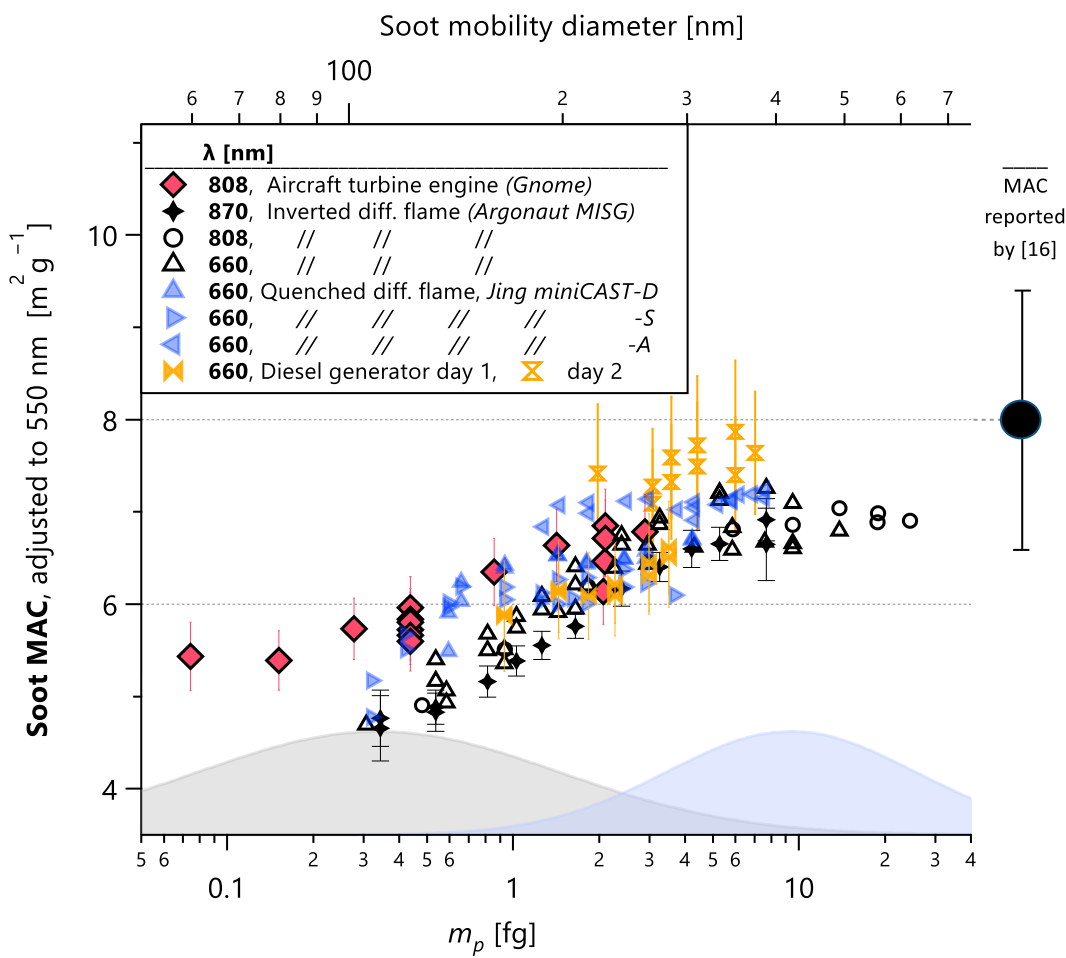

Figure 3. Size-resolved MAC for the Gnome aviation turbine engine, Diesel generator, Argonaut MISG, and the miniCAST-A, -D and -S setpoints. The measurements represent three different instruments (Table S1) adjusted to a common wavelength of 550 nm using Equation 3. $\overline{MAC_\lambda}$ is defined in Equation 2. The grey and blue shaded regions illustrate lognormal mass distributions from Table 1, with arbitrary vertical scale, of the Gnome aviation turbine engine (which produced the smallest soot particles) and the miniCAST-A setpoint (which produced large particles), respectively. For clarity, only selected error bars are shown. The upper "soot mobility diameter" axis (Eq. 6) illustrates the estimated size of the soot aggregates.

The precise influence of the size-dependent MAC on $\overline{MAC_\lambda}$ of an aerosol depends not only on the size-dependent MAC but also on the aerosol size distribution. We illustrate this in Figure 4 for the measured Gnome aviation turbine engine MACs and for a variety of GMDs. The Gnome engine MACs were parameterized using a sigmoidal fit (Equation S6) to the Gnome turbine engine data in Figure 3 (see Figure 5c for fit result). The size-integrated MAC was obtained by weighting the size-dependent MAC function by a simple lognormal soot size distribution at the GMDs given by the abscissa and the GSDs labelled in the graph. For the small soot particles from the Gnome engine, the uncertainty introduced by our simple lognormal representation is expected to be negligible [66]. The figure also plots the GMDs reported previously and discussed above.

For GMDs of 10 nm to 100 nm and reasonable GSDs of 1.6 to 2.0, $\overline{MAC_\lambda}$ spans almost the entire range measured for the size-resolved MAC. From the upper plateau to the lower plateau, the size-integrated MAC decreases 18% from the plateau value of 6.6 m² g⁻¹ to 5.4 m² g⁻¹. For the more limited range of GMDs emitted by a single engine, a MAC reduction closer to 10% might be expected. This 10% is comparable in magnitude to the 9% standard deviation of recently reported values of $\overline{MAC_\lambda}$ in the literature [16], but represents a measurement bias rather than an uncertainty. For example, to our knowledge, no previous measurements have been capable of distinguishing this change in MAC since the accepted techniques for measuring aviation turbine engine soot all measure $M_{PM}$ using techniques that are sensitive to changes in MAC. Such previous measurements therefore lacked an independent reference, such as that given by the CERMS used in our work.

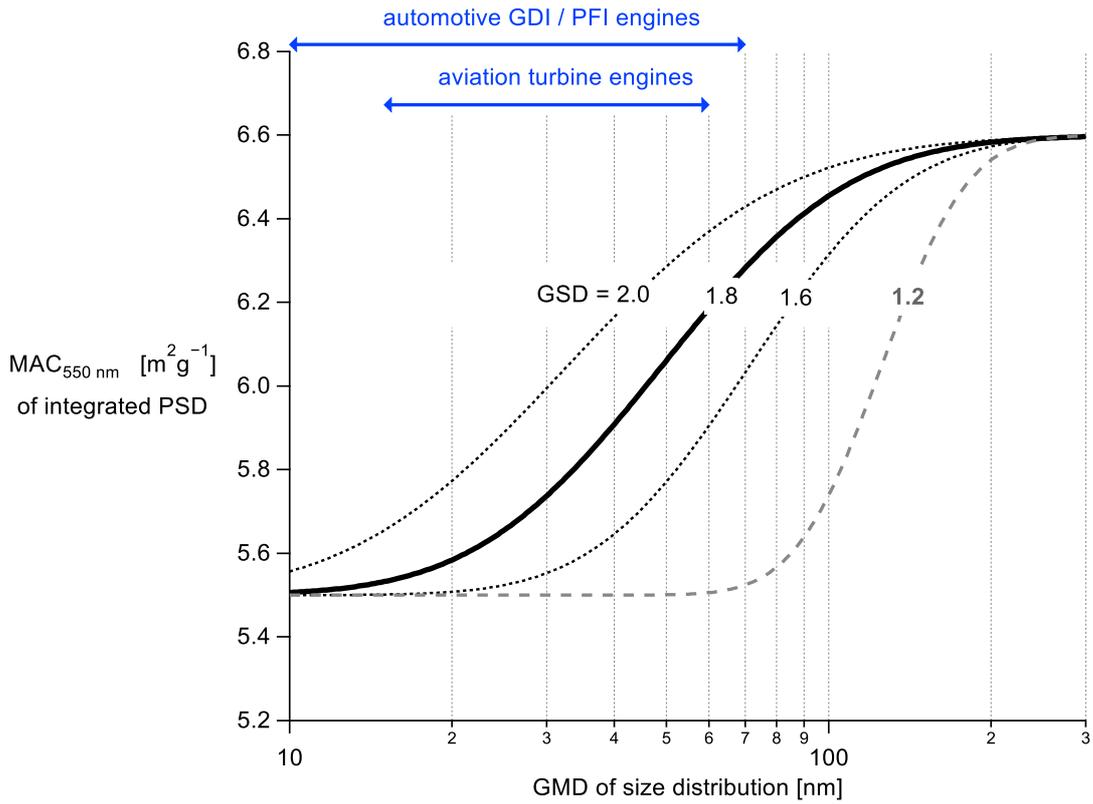

Figure 4. Size-resolved MAC from the sigmoidal fit to Gnome data in Figure 3 weighted by lognormal distributions spanning the GMD range reported previously for aviation turbine engine [64] and automotive engine [65] soot particles. The GSDs of 2.0, 1.8, and 1.6 represent plausible size distributions; the GSD of 1.2 represents the narrower distributions produced by the CPMA. If a MAC corresponding to large particles is assumed for small soot particles, the resulting positive bias in calculated mass concentration may be as high as 18%

*3.3. Literature comparison*

Figure 5 places our measurements in the context of previous size-resolved MAC measurements from diffusion flames (Figure 5a,b,d,e) and flame spray pyrolysis (Figure 5e). The trends shown by these literature data are all similar to those in our data, despite the use of various experimental techniques in each study. For example, Dastanpour et al. [27] classified particles by mass-to-charge ratio using a CPMA (similar to our work) while the other studies classified particles by mobility-to-charge ratio using differential mobility analyzers (DMAs). Other differences in particle charge, charge correction approaches, and light-absorption measurements are summarized in Table S2. This variety of experimental approaches reduces the likelihood that experimental bias caused the observed trends.

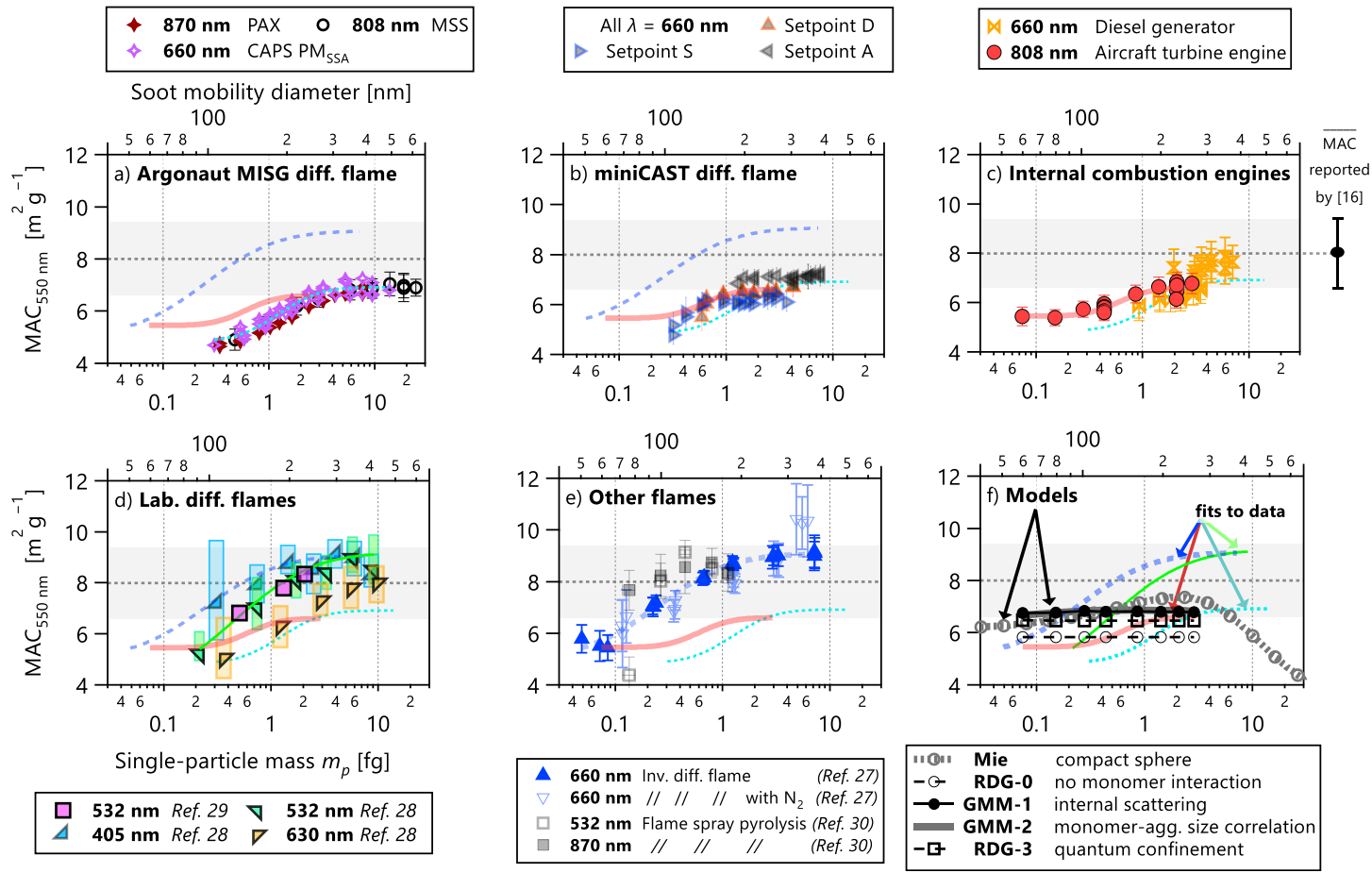

Figure 5. Size-resolved MAC for the samples measured in this study (a-c) and literature [27–30] (d-e), and modelled in this study (f). In (f), Model RDG-0 represents the null hypothesis; Models GMM-1, GMM-2, and RDG-3 represent the hypothetical causes of size-dependent MAC discussed in the text; and the (unphysical) Mie model is included for context. Grey shading shows the mean $MAC_{550nm}$ and $k = 2$ uncertainty reported by recent size-integrated studies [16]. All fits are empirical and are reproduced in (f), some fits are reproduced in (a)-(e) to facilitate comparison. The vertical bars in (d) represent upper and lower quartiles of the compiled data set of Forestieri et al. [28]; all other error bars represent standard errors. All data have been harmonized to λ = 550 nm assuming AAE = 1.

All of the studies reported in the literature may be described by the sigmoidal function mentioned above, although most studies do not report a broad enough range of sizes to differentiate between this sigmoidal function and simpler alternatives (e.g. quadratic or power-law). In particular, the study by Khalizov et al. [29] measured only three different particle sizes and the study by Kholghy and DeRosa [30] measured five.

Dastanpour et al. [27] specifically discussed a size-dependent MAC for inverted-diffusion-flame soot and postulated a power-law fit of MAC to $m_p$. Our new data and our compilation of literature data clearly show a power-law fit (which would appear as a straight line on our semi-log plots) poorly describes the overall trends for the size-dependent MAC of soot. We also note that Dastanpour et al.'s fit was strongly influenced by the high outliers (light-blue inverted triangles shown in Figure 5e) which were measured when the fuel was mixed with $N_2$, uniquely different from all other measurements plotted. We excluded this $N_2$-diluted-flame soot data set and plotted a new, purely empirical fit to the data (blue triangles shown in Figure 5e) using the same sigmoidal function as the other data sets (Equation S6, dashed blue line in Figure 5e), which is consistent with the trend for those data sets.

The study of Forestieri et al. [28] is also consistent with our interpretation. That study compiled data from four separate studies performed over the course of 7 years; we therefore plotted bars (upper and lower quartiles, with symbols for the median) rather than points in Figure 5d. Nevertheless, their results, representing measurements at 405 nm, 532 nm, and 630 nm, are fully consistent with the discussion above. While there is some apparent wavelength dependence to the trend, it is secondary to the MAC–size trend. We note that a large fraction of the data reported by Forestieri et al. [28] was produced by an inverted methane flame similar to that used by Dastanpour et al. [27]. Finally, two other studies reporting size-resolved MAC measurements by Cross et al. [32] and You et al. [31] focussed on particles with $m_p \gg 1$ fg and thus did not observe the sigmoidal trends discussed here.

The variability between our reported measurements for different soot sources is much smaller than the variability between studies of similar sources (e.g. the laboratory diffusion flames in Figure 5b and Figure 5d; and the inverted flames in Figure 5a and Figure 5e; see Figure S7 for a single plot overlaying these data). From this observation, it is inferred that the variability in calibration accuracy or experimental uncertainties is likely an important influencing factor in the MACs reported in the literature. Therefore, the most reliable estimates of the typical value of soot MAC may be from reviews of MAC measurements made in multiple laboratories, such as [16,26]. Furthermore, when normalized and overlaid (Figure 6; also shown without normalization in Figure S6), the combined measurements display variability not only in the value of MAC but also in the rate of change of MAC with size, suggesting that the observed MAC–size trends vary between sources.

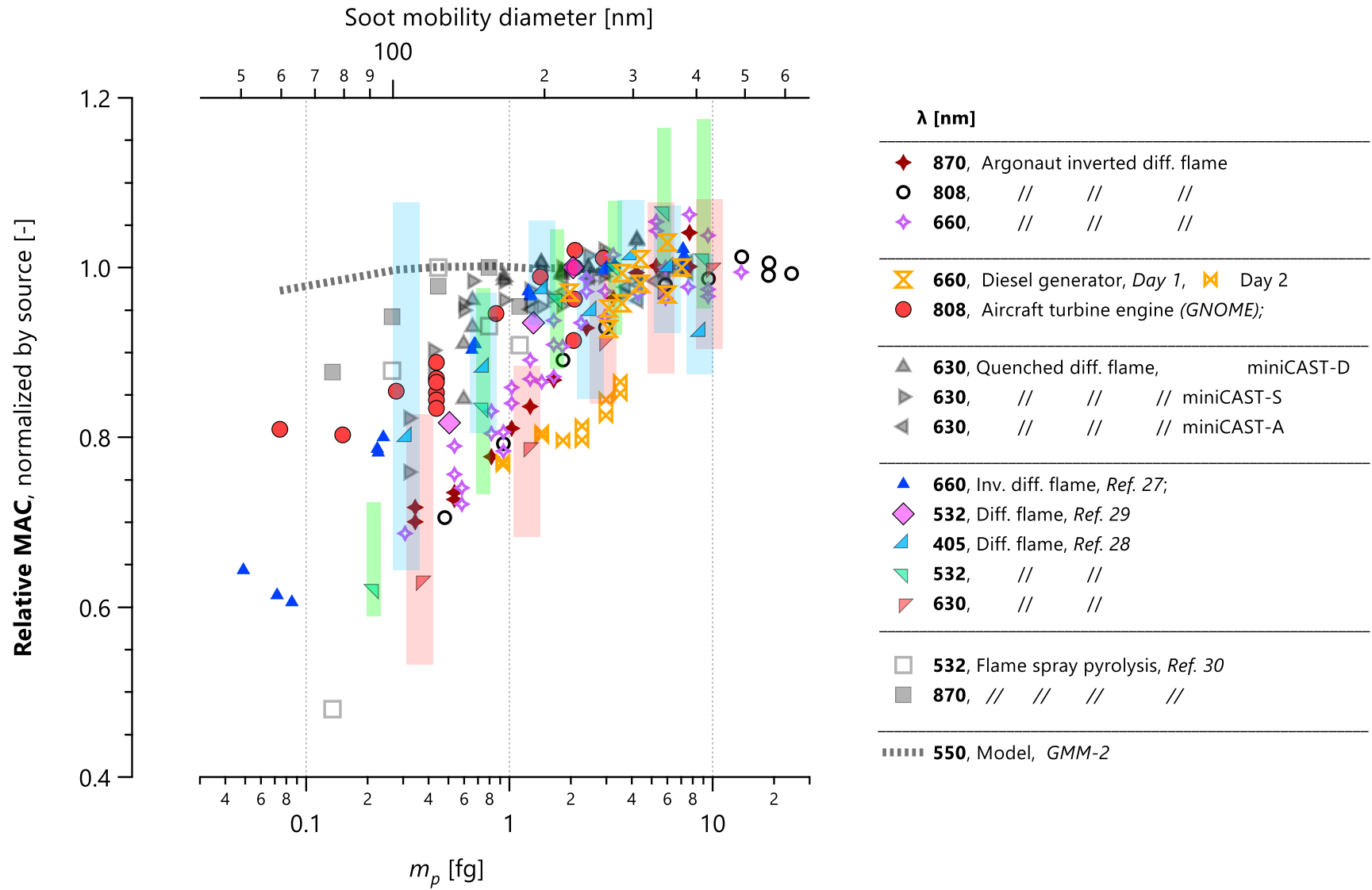

Figure 6. Similar to Figure 5 but with data overlaid on a single panel and normalized to the maximum MAC of each source. A single normalization factor was used the two diesel generator days while three individual normalization factors were used for the three miniCAST setpoints. The model GMM-2 was chosen arbitrarily for illustration.

*3.4. MAC models and hypothetical causes of size-dependent MAC*

Figure 5f shows the modelled MAC results described in Section 2.7. and Table 3. The models represent three different literature hypotheses which might produce a size-dependent soot MAC (1–aggregate internal scattering; 2–correlation between monomer and aggregate size; 3–quantum confinement effects; see Introduction). The figure also includes a null-hypothesis RDG model numbered 0 (which excludes all 3 of the above hypotheses) and a Mie model, which is used for context in the following discussion. Figure 5f is reproduced in Figure S6 with an additional axis illustrating the number of primary particles in each modelled soot aggregate.

The Mie model in Figure 5f (grey open circles and dotted line) is a simple model representing the soot aggregate as a volume-equivalent sphere. It is the only model to predict a size dependent MAC, but does not capture the MAC plateau at $m_p > 1$ fg nor the rapid decrease in MAC at smaller sizes. The limitations of Mie theory in describing soot optics are well known [26,67] and the Mie model is not included as a hypothetical explanation for the observed size-dependent MAC. Rather, the Mie model in Figure 5f was included because Forestieri et al. [28] proposed that a Mie approach may be empirically useful if the soot RI is treated as a fitting parameter at small size parameters ($x < 0.9$). Our observed variability in the MAC as a function of combustion source shows that this approach cannot accurately represent size-resolved MACs unless done on a source-specific basis.

The RDG model (black open circles and dashed line) in Figure 5f represents the simplistic assumption of non-interacting soot primary particles. This model has no size-dependent parameters and therefore shows no size dependence. This RDG model represents the null hypothesis where none of the three literature hypotheses are represented and serves as a point of reference for the other models.

In contrast to the RDG model, the GMM–DLCA model (Figure 4f, GMM-1, black filled circles and solid line), accounts for primary-particle interactions within a DLCA soot aggregate. The GMM–DLCA modelled MAC is 14% higher than the RDG-0 model ($h$ = 1.14), but also shows negligible size dependence across the size range of interest ($\Delta h$ < 0.011). Thus, there is a negligible size dependence of modelled MAC on $d_m$ in our measured range.

Our modelled $h$ can be compared to the $h$ predicted by Kelesidis and Pratsinis [39], who modelled soot particles in greater morphological detail than our model of point-contact spheres. As mentioned in the introduction, they modelled surface growth and therefore necking and overlap between monomers. They reported a parameterization of $h$ with $d_m$ describing their modelled particles. However, their parameterization also does not predict a substantial size dependence of $h$ in our measured size range; using their parameterization, we calculated $1.19 < h < 1.22$ for the Gnome aviation turbine engine soot particles. This suggests a negligible (~5%) influence of monomer overlap on the MAC of soot aggregates formed by point-contact monomers in our observed size range, consistent with the conclusions of previous modelling work [19,68].

Parameterizing the correlation of $d_{pp}$ with $d_m$ (grey line in Figure 5f) within a GMM model (Figure 5f, GMM-2, solid thick grey line) also yielded negligible size dependence across our range of interest. The parameterization changed the range of $N_{pp}$ for the Gnome aviation turbine engine data set from 9 to 332, to 19 to 159 (Figure S6). Therefore, we rejected the hypothesis that this correlation may explain the MAC size-dependence.

The final model in Figure 5f explores the influence of soot composition on MAC by parameterizing the soot RI and density following Kelesidis and Pratsinis [35], whose work built on that of Wang and coworkers [36] by considering the size-dependent optical band gap $E_g$, which was described in terms of a quantum confinement model. With this quantum confinement model (Figure 5f, RDG-3, black open square circle and dashed line), our parameterized $m$ and $\rho$ spanned narrow ranges (1.66 + 0.75$i$ to 1.66 + 0.76$i$, and 1613 to 1625 kg m$^{-3}$), so our modelled MAC spanned narrow ranges. The quantum confinement model results were not substantially altered by setting $\rho$ = 1800 kg m$^{-3}$ or by using $d_{pp}$ from Equation 11 in place of $d_m$. For the quantum confinement model to display a strongly size-dependent soot MAC, we would need to reformulate the $m = m(E_g)$ or $E_g = E_g(d_m)$ parameterizations of Kelesidis and Pratsinis [35]. As was the case with the approach of Forestieri et al. [28], the observed variability in size dependence between sources creates challenges for this approach.

Thus, none of the effects of internal scattering, monomer-aggregate size correlation, or quantum confinement, can currently explain the size dependence of MAC we have observed in this study and corroborated with the literature results. As noted in the introduction, the additional hypothesis of monomer overlap or necking has already been excluded in previous studies. A second additional hypothesis is that the soot particles studied have restructured to more compact morphologies than DLCA would predict. This hypothesis can be rejected based on the weak size dependence of our Mie model results, which represented the extreme compactness of an equivalent-volume sphere. Morphological effects are further discussed in Section S1.5, Figure S8, and Figure S9 with respect to our measurements of single-scattering albedos (SSAs).

The above process of elimination shows that the measured size-dependent MAC is not predicted by detailed optical models if size-independent physical properties of soot are assumed. Therefore, for all optical models including the RDG theory summarized in Equations 8 to 10, some physical property must vary with size. In particular, either the $E_g$ or its associated RI (Equation 10) or the density (Equation 9), or both, must vary with size. To explore this, we used Equations 8 to 10 to convert our measured MACs to $E_g$. We assumed two extreme values of density (1400 and 1800 kg m$^{-3}$) as well as a recently published parameterization [73] which interpolates between those extremes [73]. The calculated $E_g$ increases with decreasing MAC for all assumed densities (Figure S5). However, for the full range of assumed densities, no single value of $E_g$ predicts the range of observed MACs. Therefore, RDG theory requires that not only the density but also the refractive index change with size. A size-dependent refractive index implies a size-dependent $E_g$ (Equation 9), and a size-dependent $E_g$ implies a size-dependent degree of graphitization (maturity) of soot [23]. This conclusion is consistent with previous Raman spectroscopy measurements of a size-dependent soot maturity (degree of graphitization) for inverted-diffusion-flame soot [27,34].

A physical model of the size-dependent MAC will require further work to elucidate the precise size-dependence of soot $E_g$ and density. However, since Figure 5 does not show a universal size-dependence of MAC for different sources, there is unlikely to be a corresponding universal size-dependence of density and $E_g$. This lack of a universal size dependence may indicate differences in the temperature, chemistry and residence time experienced by particles in different combustion sources.

### *3.5. Implications for CERMS calibrations*

Recent studies have discussed the calibration of black-carbon or soot instruments using a laboratory soot source with an experimental configuration identical to that discussed here, and referred to as the CPMA-Electrometer Reference Mass Standard (CERMS) [47,50–52]. We have used an identical setup in our experiments. Because the CERMS always uses CPMA-classified particles, one CERMS calibration point corresponds to one mass-classified (referred to above as size-classified) condition.

A full CERMS calibration is a set of mass-classified points, each of which takes approximately 3 minutes to obtain [47]. This set of points may span several particle sizes, and may be viewed as a set of point-wise calibrations at different sizes. Our identification of a size-dependent MAC implies that these point-wise calibrations will differ systematically from one another. Such a systematic trend is not a measurement artifact, but a feature of the soot source used for calibration. Moreover, our literature review suggests that this feature is likely present in all soot sources. While calibration techniques which intrinsically integrate mass over all particle sizes will not be able to observe this feature, it would nevertheless have been present in the source particles prior to integration. In this context, the unique feature of the CERMS is its ability to obtain a size-dependent calibration.

The influence of this feature is relatively minor if calibration is performed at a range of CPMA setpoints. For example, if the Gnome aviation turbine engine data presented in Figure 3 are treated as calibration data (all sizes), the calibration residuals are only scattered about the regression line by < 10% (Figure S10) and the uncertainty in a weighted linear regression is < 4.1 %. This scatter is an extreme example, since in practice, a CERMS calibration should not equally weight data taken at the tails of the size distribution.

## 4. Conclusions

We measured the relationship between MAC and particle mass for soot from four different combustion sources. With decreasing soot particle mass, we observed that the MAC decreased from an upper plateau of approximately 8 $m^2 g^{-1}$ at roughly 4 fg (about 300 nm mobility diameter) to a minimum value of at least 4 $m^2 g^{-1}$ at smaller masses/sizes. When integrating over entire soot size distributions, this range becomes narrower (Figure 4). The general trends we observed are consistent with previous literature measurements, but the size-dependent MAC appears to vary between combustion sources. Taken together with the literature measurements, the data suggest that the decrease from the MAC upper plateau may vary in its starting point and rate of decrease between sources.

When the size-dependent (or, equivalently, mass-dependent) MAC is weighted by the soot mass distribution (cf. shaded curves in Figure 3), the resulting average MAC will be affected most strongly by the small particles produced by internal combustion engines such as the Gnome aviation turbine engine measured here. Aviation turbine engines and automotive GDI and PFI engines produce significantly smaller soot particles than other sources, with GMDs in the range 10 nm to 70 nm, and therefore the smallest and largest particles may have substantially different MACs. Changes in the emitted particle size distributions may potentially correspond to changes in $\overline{MAC_\lambda}$ and consequently to significant biases in the reported BC or nvPM mass concentrations of absorption-based instruments.

We modelled soot MAC using different methods and various hypotheses from the literature that could potentially explain the size-dependent MAC of soot: aggregation effects ($h$), monomer-aggregate size correlation ($d_{pp} \propto d_m$), and changing composition ($E_g \propto d_m$ and $\rho \propto d_m$ according to literature parameterizations). None of these hypotheses can reproduce the observed size-dependent MAC. To predict the observations, RDG theory requires a size-dependent $E_g$ as well as a possible size-dependent density, which implies a size-dependent soot maturity. Moreover, the observed size-dependence varied between combustion sources, suggesting that the temperature, chemistry and residence time experienced by particles before emission differ between sources. Future work should seek to parameterize these differences in terms of easily measured flame conditions.

## 5. Acknowledgement

Technical support from Daniel Clavel and Simon-Alexandre Lussier (in situ measurements) and Brett Smith (microscopy images) is greatly appreciated. We thank the Rolls-Royce team in Derby, UK, for their support. We are grateful to Andrew Crayford at Cardiff University, UK, for providing the MSS; Hongsheng Guo and Shouvik

## 6. Author contributions

Conceptualization – JCC, GJS, PL, FL; Investigation – JCC, GJS, MPJ, PL; Formal analysis – JCC, TJJ, FL, TAS; Methodology – JCC, GJS, TJJ, FL,MPJ, PL; Writing original draft – JCC; Writing review and editing – all authors.

# S1. Supplementary information

This supplementary information contains:

- Additional text on data analysis

- Additional tables from our literature review of MAC measurements

- Additional visualizations of our analysis and results

## *S1.1. Additional details on average UDAC-CPMA single-particle mass calculation*

The conceptual basis for the average UDAC-CPMA single-particle mass calculations was outlined in Section 2.6. Here we provide some additional discussion of both approaches.

The iterative-average-charge (IAC) approach (Figure 2a) estimated the average single-particle mass classified by the CPMA, $\overline{m_p}$ or simply $m_p$, by iteratively solving

$$\overline{m_{p,k+1}} = \overline{m_{p,k}} \times \bar{q} = \overline{m_{p,k}} \times q_F(d_C[\overline{m_{p,k}}]) \qquad \text{S1}$$

Where $\overline{m_{p,k}}$ and $\overline{m_{p,k+1}}$ are initial and improved estimates of $m_p$ and $\bar{q}$ is the average charge for a particle of mass $\overline{m_{p,k}}$ predicted using the Fuchs charging model [60]. The Fuchs model requires a diameter rather than a mass as input; this charging diameter $d_C$ was assumed to be equivalent to $d_m$, with $d_m$ estimated from $\overline{m_{p,k}}$ using the effective density as described in Section 2.6. The initial guesss of $\overline{m_{p,k}}$ is estimated from the CPMA setpoint *m/q* by assuming *q*=1, and the equation is iterated until the solution converges.

Under our conditions[1], results of the Fuchs charging model were adequately represented by a curve with the parameterization of

$$\bar{q} = 6.4168 \cdot m_p^{0.4204} + 0.4587 \qquad \text{S2}$$

Where $\bar{q}$ is the average number of charges on a particle of mass $m_p$ (in fg) charged by the UDAC at an ion·charge product of $10^{13}$ ions·s/m³. This parameterization is useful given the computational complexity of the Fuchs charging model.

The mass-and-charge distribution approach (Figure 2b) was based on the mass distribution of the classified particles estimated by the CPMA transfer function as follows. For each soot source, we represented the particle population using lognormal input distributions in mobility diameter $d_m$ ($dN/dlogd_m$). These distributions were described by the measured geometric count mean diameters CMD and geometric standard deviations GSD of each source (Table 1). The charge distribution $q(d_m)$ on these particles after the UDAC, at integer multiples of $m/q$, was predicted using the Fuchs model [60] using a dielectric constant $\kappa$ of 13.5. This dielectric constant is an average of the range used by Ouf and Sillon [70] for soot, and our results were relatively insensitive to this assumption. The lognormal distributions were then converted to mass distributions ($dN_{in}/dlogm_p$) based on Equation 7. The transport of this mass distribution through the CPMA was predicted using a CPMA transfer function calculated for particles with $q = 1, 2, 3, ...$ for all cases where $dN_{in}/dlogm_p$ was substantially greater than zero. The CPMA transfer functions were calculated using the finite-difference method outlined by Sipkens et al. [57]. The overall CPMA output mass distribution $dN_{out}/dlogm_p$ was then calculated by combining the input mass distribution wth the charge distribution and transfer function. This calculation assumes that the charge distribution is approximately constant across the narrow width of the CPMA transfer function for each particle charge state.

---

[1] Aerosol pressure and temperature of 101,325 Pa and 298.15 K, respectively, as well as assuming the soot particles follow the mass-mobility relationship established by Olfert and Rogak [58].

The $dN_{out}/dlogm_p$ predicted by the mass-and-charge distribution approach can be used to estimate $m_p$ directly. Since $m_p$ is the average mass of the classified particles, its value is

$$m_p = \frac{1}{N_{\text{out}}} \sum_j m_j \cdot dN_{\text{out},j} \qquad \text{S3}$$

Where $dN_{out,j}$ and $m_j$ are the number and single-particle mass of particles with classified by the CPMA, respectively. These quantities are summed for all bins in the $m_p$ distribution (i.e. integrated over single-particle mass). We then attributed the MAC calculated with Equation 1 (main manuscript) to particles of mass $m_p$. This average-mass approach, implicitly assumes that the mass distribution of the CPMA classified particles is narrow[2] (such that most of the particles contributing to $m_p$ have similar mass) and that the MAC of the individual particles is constant over this narrow distribution.

As mentioned in Section 2.6. , we also used the $dN_{out}/dlogm_p$ predictions to consider the possibility that particles of different $m_p$ and therefore different MAC were transmitted by the CPMA at a single setpoint. To this end, we used Equation 1 to describe each unique particle mass $m_j$ as having its own MAC, $\text{MAC}_{m_j}$ and its own mass loading $M_{PM,j} = N_j m_j$:

$$\text{MAC}_{m_j} = \frac{B_{\text{abn},j}}{N_j m_j} \qquad \text{S4}$$

---

[2] The width of the classified mass distribution is a function of the combined widths of the particle charge distribution, CPMA transfer function and polydispersed aerosol.

Where $N_j$ is the total number concentration of particles with mass $m_j$ and charge j. Note that a particle with mass $m_j$ might be transmitted by the CPMA as a singly charged particle, doubly charged particle, or other multiply charged particle $(m_j = m_1/1 = 2m_2/2 = qm_3/q)$. This fact can be captured using a system of equations and simultaneously solved to determine $MAC_q$ for all $m_q$.

In order to construct a system of equations that is over-constrained (more measurement points than degrees of freedom), we first consolidated the mass fractions of the CPMA classified particles ($f_{i,j} = n_{i,j}m_j / \sum_j n_{i,j}m_j$) into a matrix of common mass bins (columns) for all test points (rows), then consolidated this matrix into a number of log-spaced mass bins that was a factor of 2 to 4 smaller than the number of test points. The consolidated matrix was further simplified by (a) removing mass bins (i.e. columns of matrix) that were often empty (i.e. $f_{i,j} > 0.01$ in fewer than three bins at a given value of $j$) and (b) by removing test points where the mass distribution was no longer adequately represented (i.e. $\sum_j f_{i,j} < 0.99$) after this consolidation process. The second criteria was rarely implemented and was mostly applied to samples that had multiple mass bins removed for not meeting criterion (a) due to not overlapping in the mass domain of at least two other samples.

The consolidated matrix can be written by defining $f_{i,j}$ as the mass fraction of particles with mass $m_j$ at test point $i$ as:

$$\begin{bmatrix} f_{1,1} & \cdots & f_{1,j} \\ \vdots & \ddots & \vdots \\ f_{i,1} & \cdots & f_{i,j} \end{bmatrix} \begin{bmatrix} MAC_{m_1} \\ \vdots \\ MAC_{m_j} \end{bmatrix} = \begin{bmatrix} MAC_1 \\ \vdots \\ MAC_i \end{bmatrix} \qquad S5$$

which is a system of equations that represents the measured $MAC_i$ as a sum of the (potentially) different $MAC_{m_j}$ for each single-particle mass bin $m_j$. This system of equations was solved by applying least-squares minimization inversely weighted by the standard deviation of the average measured MAC at each test point plus 1% of the average measured MAC. The 1% offset was included in the fit normalization to reduce the preferential weighting that least-squares minimization inherently places on the largest values measured and follows previous recommendations [71]. This second method based on the estimated mass distribution is referred to as the system-of-equations approach, and its uncertainties for a k=1 coverage factor were determined by bootstrapping (repeatedly and randomly sampling a subset of the complete data set).

*S1.2. Validation of iterative-average-charge (IAC) approach to $m_p$*

As mentioned in Section 2.6., we also used the $dN_{out}/dlogm_p$ predicted by the mass-and-charge distribution approach (Figure 2b, Section 2.6., and preceding SI subsection) to validate the simpler IAC algorithm by calculating the relative number of particles of each charge state *q* exiting the CPMA at a given setpoint.

Figure S2 shows that the three different approaches produce similar results (iterative average charge, average single-particle mass or system-of-equations approaches). The figure uses data from the CAPS PM$_{SSA}$ monitor measuring MISG soot as an example.

The IAC approach agreed closely with the average-single-particle-mass approach for all test points, despite its implicit assumptions that the sampled particle size distribution and the CPMA transfer function may be neglected. This unexpected agreement can be understood by reference to the more detailed calculation of the $m_p$ and charge distribution exiting the UDAC–CPMA. The distribution of total charge as a function of $m_p$ was approximately lognormal (Figure S1 and Figure S3) due to the large number of charges per particle imparted by the unipolar charger in the UDAC. Our observed agreement implies that this charge distribution played the major role in determining the average $m_p$, while the sampled particle size distribution and CPMA transfer function played minor roles. At present, this conclusion is specific to the conditions of our study; further investigation is required to validate these implications and identify limitations of the IAC approach.

Finally, the system-of-equations approach showed a similar trend as the average-single-particle-mass approach in Figure S2. This agreement confirms that the average-mass approach, and its associated assumptions (as previously described) were valid for the data collected in this study. In the following discussion, we focus on the results produced by the average-mass approach because the system-of-equations approach involved grouping the data in the $m_p$ dimension in order produce an over-constrained system where the system-of-equations could be fitted successfully.

*S1.3. Uncertainty in iterative-average-charge (IAC) approach*

The preceding subsection showed that the assumptions behind the IAC approach (Figure 2a) impart minor uncertainty to the result, compared to a complete calculation considering aerosol mass and charge distributions (Figure S2). In this section, we discuss two types of uncertainties which remain in the approach.

The first uncertainty is due to the propagation of uncertainties through the mass-mobility relationship (Equation 6 and 7 in the manuscript). We quantified this uncertainty by a perturbation analysis whereby the inputs to the mass-mobility relationship were perturbed systematically (Table S3). Two of these, $\rho_{\text{eff},100}$ and $\varepsilon$ originate in the effective density power law (mathematically equivalent to the mass-mobility relationship) while the third, $n_i t$, originates in the UDAC setpoint. For simplicity, we omitted a perturbation of the dielectric constant because it is a multiplicative factor in predicting $q$ [72], so would have the same effect as perturbing $n_i t$. As expected from an inspection of Equation 6 or 7, Table S3 shows that the predicted $m_p$ is most sensitive to the mass-mobility exponent $\varepsilon$, with a $\pm\,5\%$ change in $\varepsilon$ leading to a $+39\%/-24\%$ change in $m_p$ at the extreme upper condition of $m_p = 10$ fg ($d_m \approx 500$ nm). At the more moderate condition of $m_p = 1$ fg, a $\pm 5\%$ change in $\varepsilon$ leads to a $+18\%/-13\%$ change in $m_p$.

The second uncertainty is due to uncertainty in particle morphology. Particle morphology affects the mass-mobility relationship. In the most extreme case, open DLCA aggregate structures ($\varepsilon = 2.48$) become quasi-spherical ($\varepsilon \approx 3$, a $+17\%$ change) after the condensation of volatile materials. A correlated change in $\rho_{\text{eff},100}$ is expected and must be considered. Considering the extreme case of highly-compact particles based on measurements made in our laboratory (for which $\varepsilon = 2.97$ and $\rho_{\text{eff},100} = 411.35$ according to a fit to Equation 6; where the value of $\varepsilon$ close to 3 indicates near sphericity), we calculated a difference in $m_p$ of 11% and 39% at 110 nm and 240 nm respectively. In other words, when soot particles are assumed to be DLCA aggregates, their true mass may be 39% higher at 240 nm than predicted. This corresponds to a shift by a factor of 1.39 on the abscissas of Figure 3, Figure 4, and Figure 5, which is negligible considering the two orders of magnitude spanned by those axes.

Morphology also affects particle charging. For particles of $d_m = 100$ nm, agglomerates charged by unipolar diffusion chargers have < 20% more charge-per-particle than spheres of the same $d_m$ [59,72]. A more precise quantification of this charging uncertainty is difficult due to the complexity of charging models, as well as the difficulty of obtaining reliable reference data [73]. Regardless, this charging uncertainty is smaller than the morphology-related uncertainty in the mass-mobility relationship.

### S1.4. *Fit function applied to size-dependent MACs*

In Figure 4 and Figure 5, we used the error function to describe our MAC observations, with lower limit *a,* upper limit *b,* and abscissa-offset and scaling parameters *c* and *d*:

$$\text{MAC} = a + (b-a)\left[\frac{1}{2} + \frac{1}{2}\text{erf}(c \log m_p - d)\right] \qquad \text{S6}$$

We hypothesize that $a \approx 4 \text{ m}^2\text{g}^{-1}$ (the MAC of incipient soot particles described by Wan et al. [36]), $b \approx 8 \pm 0.7 \text{ m}^2\text{g}^{-1}$ (the MAC of mature soot reviewed by Liu et al. [16]) and that $c$ and $d$ vary slightly according to the discussion in the main text. However, the available data did not allow this hypothesis to be rigorously tested and all reported fits were performed without constraint.

## S1.5. Single-scattering albedo (SSA)

Our CAPS PM$_{SSA}$ measurements provide single-scattering albedos (SSAs) at 660 nm for each size-resolved MAC test point. Figure S8 and Figure S9 show these SSAs plotted as a function of average single-particle mass $m_p$ and as a function of MAC, respectively. In Figure S8, the SSA generally increases slightly from 0.12 to 0.22 as $m_p$ increases from 0.3 fg to 10 fg. The SSA trend is not clearly sigmoidal as was the case for MAC, potentially because the SSA is influenced more strongly by morphology than the MAC [18,20]. Also, the SSA trend shows less source-to-source variability than the corresponding trend of MAC versus $m_p$ (as discussed in the main text), potentially because the SSA trend reflects increasing particle size parameter $x$ rather than a changing MAC [17]. Notably, the trend for the Argonaut MISG and the diesel engine soot indicate higher SSAs for larger particles than the other sources. This may reflect more-compact soot particles being emitted from these sources relative to the others.

In Figure S9, the 660 nm SSA is plotted against the MAC reported in the manuscript. No clear trend is seen in the data. The SSA was more strongly associated with particle size than with the measured MAC.

## S1.6. *Supplementary figures and tables*

Table S1. Instruments used to measure in situ aerosol light absorption in this work. CAPS PM$_{SSA}$: Cavity attenuation phase shift–particulate matter single scattering albedo; MSS: Micro soot sensor plus; PAX: photoacoustic extinctiometer.

| Instrument | $\lambda$ [nm] | Measurement principle | Calibration source |
|---|---|---|---|
| **CAPS PM$_{SSA}$** | 660 | Extinction minus scattering | $(NH_4)_2SO_4$ |
| **MSS** | 808 | Photoacoustic | MISG soot |
| **PAX** | 870 | Photoacoustic | Graphitic nanoparticles and $(NH_4)_2SO_4$ |

Table S2. Experimental conditions of single-particle-mass- or size-resolved MAC measurements in the literature. The APM (Aerosol Particle Mass Analyzer) in this table is analogous to the CPMA. The expression "DMA–APM+CPC" (or "DMA-CPMA+CPA") describes the use of a DMA to size-classify particles that are then counted by a CPC, then multiplied by the average single-particle mass measured by an APM–CPC (or CPMA-CPC) experiment, which is sometimes called an effective density measurement.

| Study | Particle charging | Particle classification | Absorption measurement, $B_{abn}$ | Mass concentration measurement, $M_{PM}$ | Multiple charging addressed by |
|---|---|---|---|---|---|
| **This work** | Highly charged (UDAC) | CPMA | Photoacoustic (2 instruments) and extinction-minus-scattering (CAPS PM$_{SSA}$) | CPMA–electrometer | Average single-particle mass from UDAC-CPMA; checked against downstream SMPS measurement |
| **Khalizov 2009 [26]** | Equilibrium ($^{210}$Po) | DMA–DMA | Extinction-minus-scattering (house-made) | DMA–APM+CPC | Treated as negligible after two charger-DMAs |
| **Dastanpour 2017 [24]** | Highly charged (UDAC) | CPMA | Extinction-minus-scattering (CAPS PM$_{SSA}$) | CPMA–electrometer | Reference to downstream SMPS measurement |
| **Forestieri 2018 [25]** | Equilibrium (X-ray) | DMA and DMA–CPMA | Photoacoustic (3 instruments) and extinction-minus-scattering (CAPS PM$_{SSA}$) | DMA–CPMA+CPC | Measurement of multiple charges by SMPS or single-particle soot photometer |
| **Kholghy and DeRosa 2020 [27]** | Equilibrium (X-ray) | DMA | Photoacoustic (DMT PAX) | DMA–APM+CPC | Treated as negligible after one charger-DMA |

Table S3. Sensitivity of the iterative-average-charge calculation to the input parameters. The third column shows the change in $m_p$ following a perturbation of the input parameters by 5%. The fourth column shows relative sensitivity coefficients (RSCs), which are similar to the third column but normalized to give the percent change in the particle mass for a 1% increase in the input parameters. A negative RSCs indicates a decrease in $m_p$ for an increase in the given input parameter. For these calculations a UDAC $n_i t$ of $10^{13}$ ion·s/m³ was used.

| CPMA $m/q$ | Input parameter, $x$[a] | $\Delta m_p$ [%] for a +//−5% change in $x$ | RSC ($\Delta m_p$ [%] for a +1% change in $x$) |
|---|---|---|---|
| 0.1 fg | $\rho_{eff,100}$ | +3.6 / −3.3 | −1.38 |
| | $\varepsilon$ | +2.6 / −2.1 | −0.95 |
| | $\kappa$ | −0.2 / +0.2 | +0.07 |
| | $n_i t$ | −3.0 / +2.4 | +1.09 |
| 1 fg | $\rho_{eff,100}$ | +3.5 / −3.3 | −1.36 |
| | $\varepsilon$ | +18 / −13 | −6.36 |
| | $\kappa$ | −0.1 / +0.1 | +0.04 |
| | $n_i t$ | −2.9 / +2.3 | +1.04 |
| 10 fg | $\rho_{eff,100}$ | +3.5 / −3.2 | −1.34 |
| | $\varepsilon$ | +36 / −23 | −11.5 |
| | $\kappa$ | −0.04 / +0.04 | +0.01 |
| | $n_i t$ | −2.9 / +2.3 | 1.05 |

[a]Terms are $\rho_{eff,100}$: effective density of a $d_m = 100$ nm particle; $\varepsilon$: mass-mobility exponent; $\kappa$: dielectric constant; $n_i t$: ion-concentration-time product of the UDAC;

Table S4. Results of fitting Equation S6 to the data in Figure 5 (and Figure S7). n.f. = not fitted. Fits were not performed when the range of data was too narrow to constrain a sigmoidal trend.

| Source | a | b | c | d |
|---|---|---|---|---|
| Gnome turbine[a] | 5.5 ± 0.2 | 6.6 ± 0.2 | 2.9 ± 2.6 | −0.64 ± 0.91 |
| Diesel generator | n.f. | n.f. | n.f. | n.f. |
| Argonaut MISG[b] | 4.8 ± 0.5 | 6.9 ± 0.1 | 2.25 ± 0.78 | 0.21 ± 0.31 |
| miniCAST-S | n.f. | n.f. | n.f. | n.f. |
| miniCAST-D | n.f. | n.f. | n.f. | n.f. |
| miniCAST-A | n.f. | n.f. | n.f. | n.f. |
| Laboratory flame[c], Ref. [27] | 5.1 ± 0.3 | 9.1 ± 0.4 | 1.3 ± 0.2 | −0.79 ± 0.08 |
| Laboratory flame, Ref. [28] at 405 nm | n.f. | n.f. | n.f. | n.f. |
| Laboratory flame Ref. [28] at 532 nm | 4.8 ± 3.2 | 9.0 ± 0.5 | 1.7 ± 2.4 | −0.36 ± 0.63 |
| Laboratory flame Ref. [28] at 630 nm | n.f. | n.f. | n.f. | n.f. |
| Laboratory flame, Ref. [29] | n.f. | n.f. | n.f. | n.f. |
| Laboratory flame Ref. [30] | n.f. | n.f. | n.f. | n.f. |

Fitted over single-particle mass ranges of (a) 0.07 to 2.9 fg, (b) 0.3 to 13.9 fg, (c) 0.05 to 7.2 fg, (d) 0.2 to 9,1 fg.

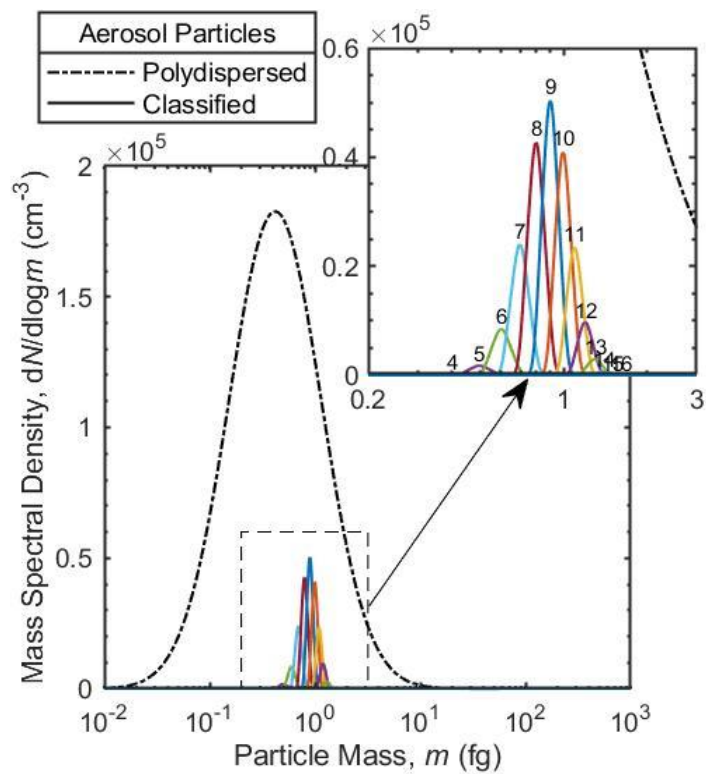

Figure S1. A representative mass distribution for particles transmitted through the CPMA after the UDAC. The example shows the input mass distribution (black line) and output mass distributions of particles with $1 < q < 15$ for miniCAST-D particles with $m/q = 0.1$ fg/e at a CPMA resolution of 3 and UDAC setpoint of $4.2 \times 10^{13}$ ions·s/m$^3$. The sum of all outputs may be approximated as a lognormal distribution.

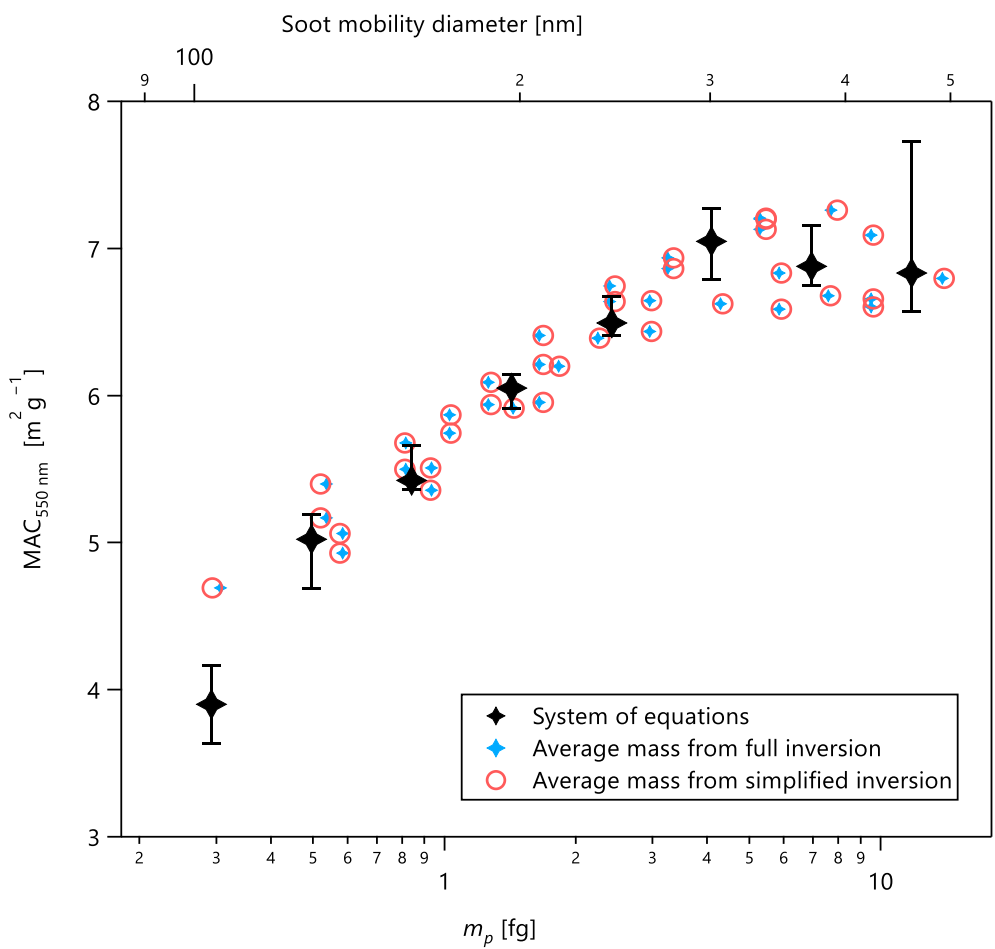

Figure S2. MAC$_{550nm}$ calculated using the three average-single-particle-mass approaches for the CAPS PM$_{SSA}$ data measured at 660 nm. Some error bars are omitted for clarity. Fewer data points are available through the system of equations for the statistical reasons described in the text. Therefore, other figures in this work used the second data set (average mass from charge-and-mass distributions, with minimal assumptions).

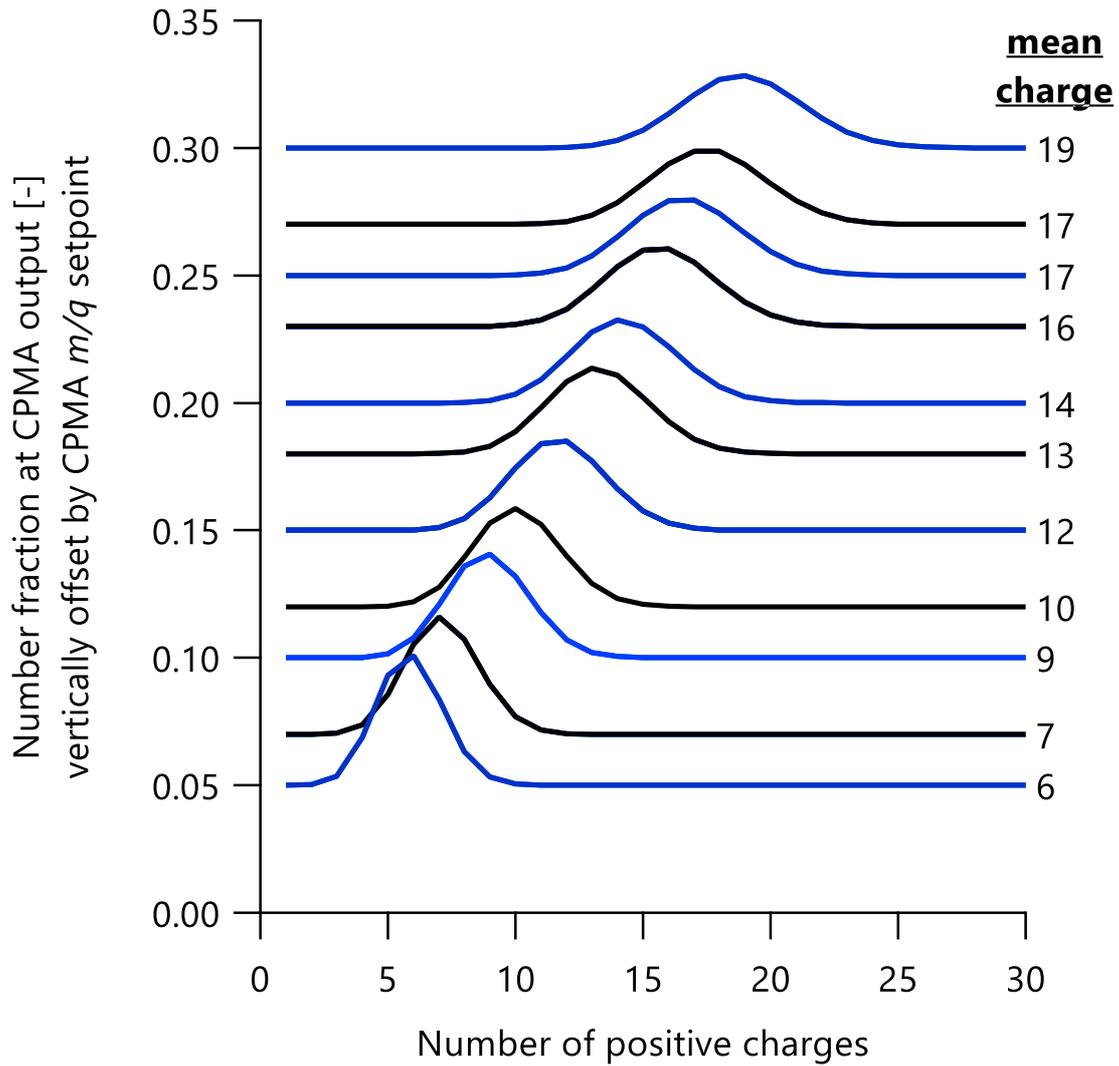

Figure S3. Similar to Figure S1 but for all CPMA setpoints in the Diesel generator experiment. The CPMA setpoint is illustrated by the intercept of the curves on the ordinate axis and is between 0.05 and 0.30 fg/e. The probability distribution of charges is shown by each curve. The mean charge is shown by the values on the right.

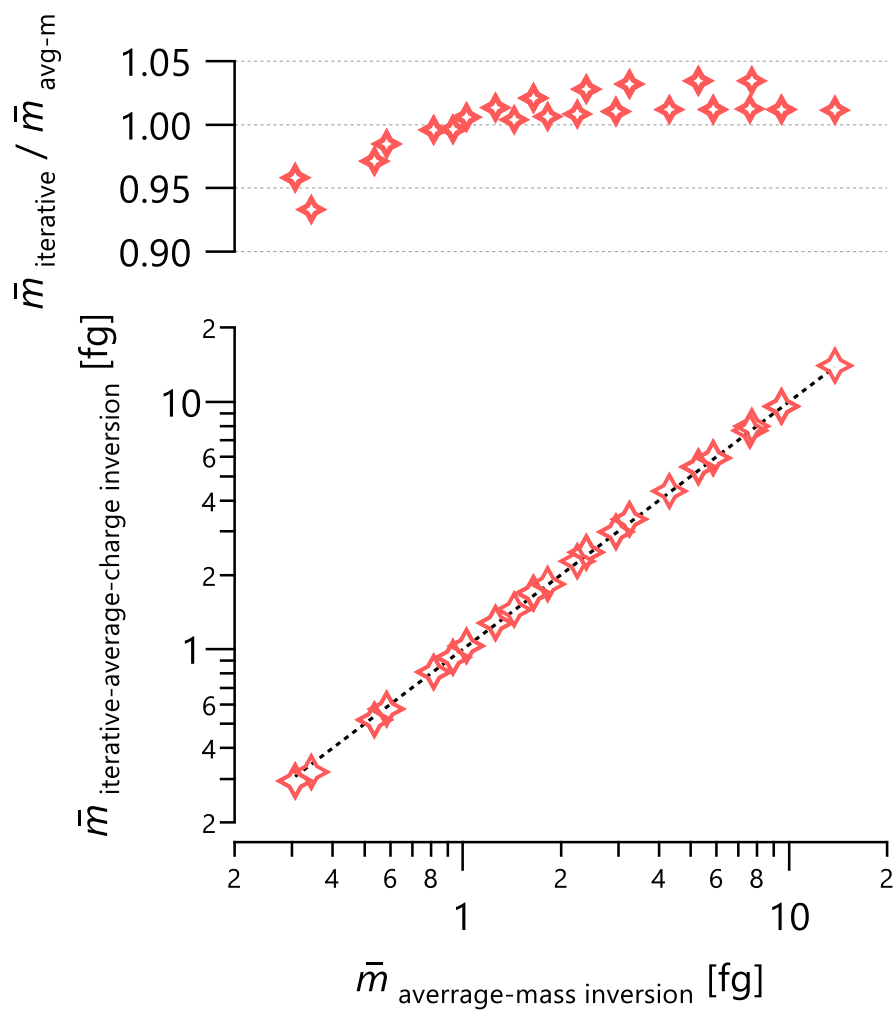

Figure S4. Comparison of the mean CPMA-transmitted single-particle mass from the iterative-average-charge (IAC) method and the average single-particle mass methods.

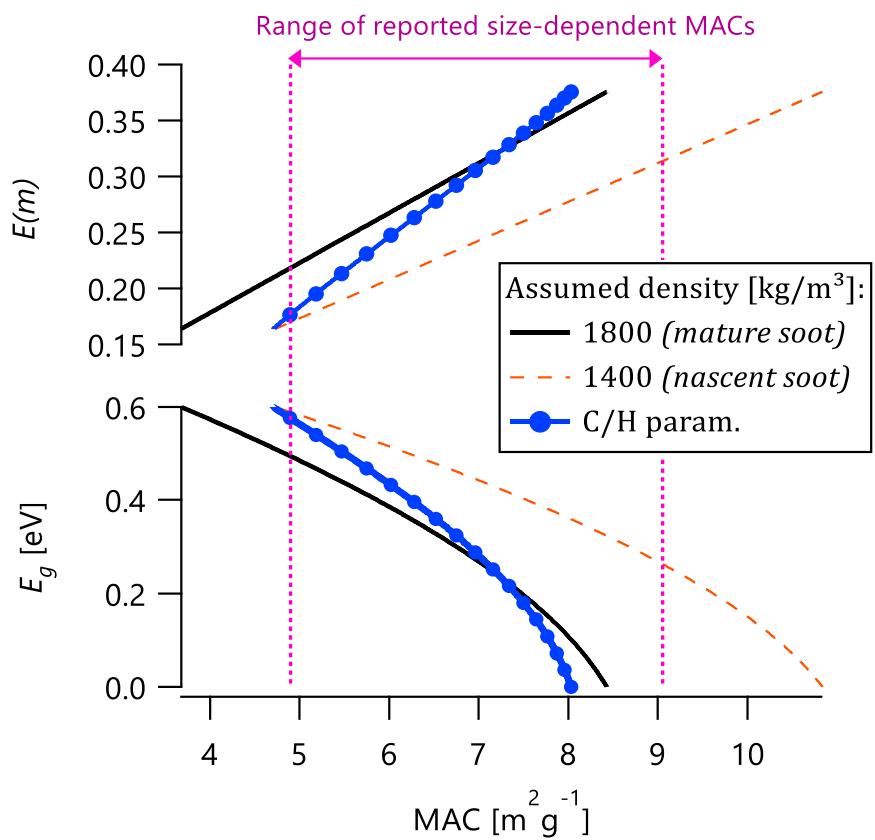

Figure S5. Band-gap energies $E_g$ (lower panel) and absorption functions $E(m)$ (upper panel) corresponding to the measurements discussed in the main text. The vertical pink lines highlight the range of reported size-dependent MACs shown in Figure 5. All curves were calculated with Equations 8, 9, and 10, similar to Ref. [39], RI $1.66 + 0.76i$, and $h = 1.14$ as calculated from the GMM-1 and RDG-0 models (main text). The three sets of curves represent different assumptions about particle density in Equation 8. In particular, the curves assume the density of mature soot, nascent (young) soot, or a maturity-dependent density based on the carbon/hydrogen-ratio (C/H) and parameterized by [74]. Although the parameterization of [74] assumed $E_g = 0.25$ eV for mature soot, we have extrapolated down to 0 eV for illustration. This parameterization may not be applicable for all flames discussed in the main text, and it does not predict the upper range of MACs reported in the main text.

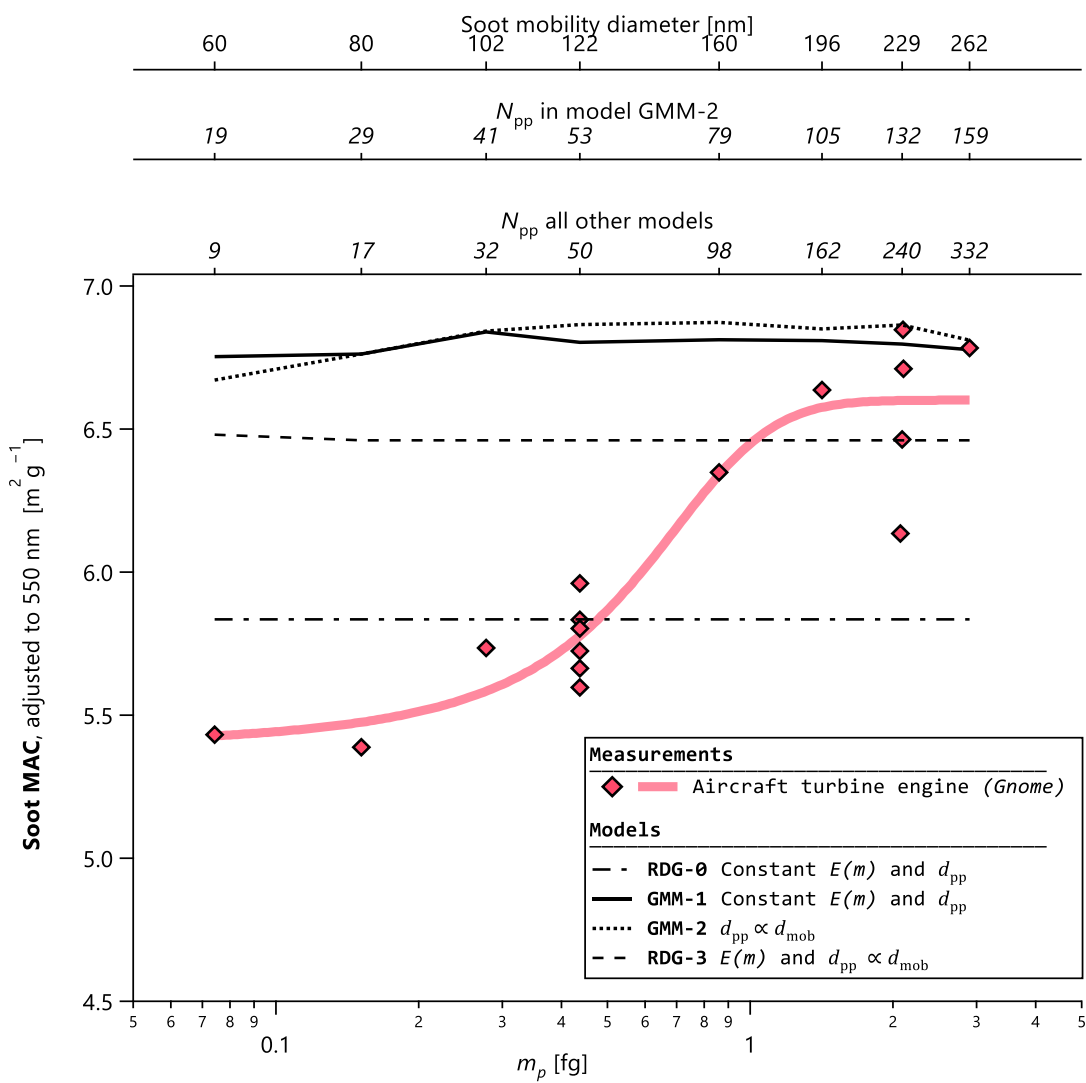

Figure S6. Measured and modelled size-resolved MACs for the Gnome aviation turbine engine. The fit to the measurements highlights the observed sigmoidal trend. We modelled 3 hypotheses for size-dependent soot properties (GMM-1, GMM-2, and RDG-3). The null-hypothesis model (RDG-0) assumes DLCA aggregates of constant $d_{pp}$ and RI. The GMM-1 model accounts for internal scattering within the soot aggregate. The GMM-2 model implements the correlation of $d_{pp}$ with $d_m$ (and $m_p$) observed previously [27,58,63]. The graphitic-shell model extends the correlation model to hypothesize that the extreme curvature of smaller soot particles makes them less graphitic. Finally, the GMM-3 model represents quantum confinement following Ref. [35]; it does not represent the $d_{pp}$ of this engine.

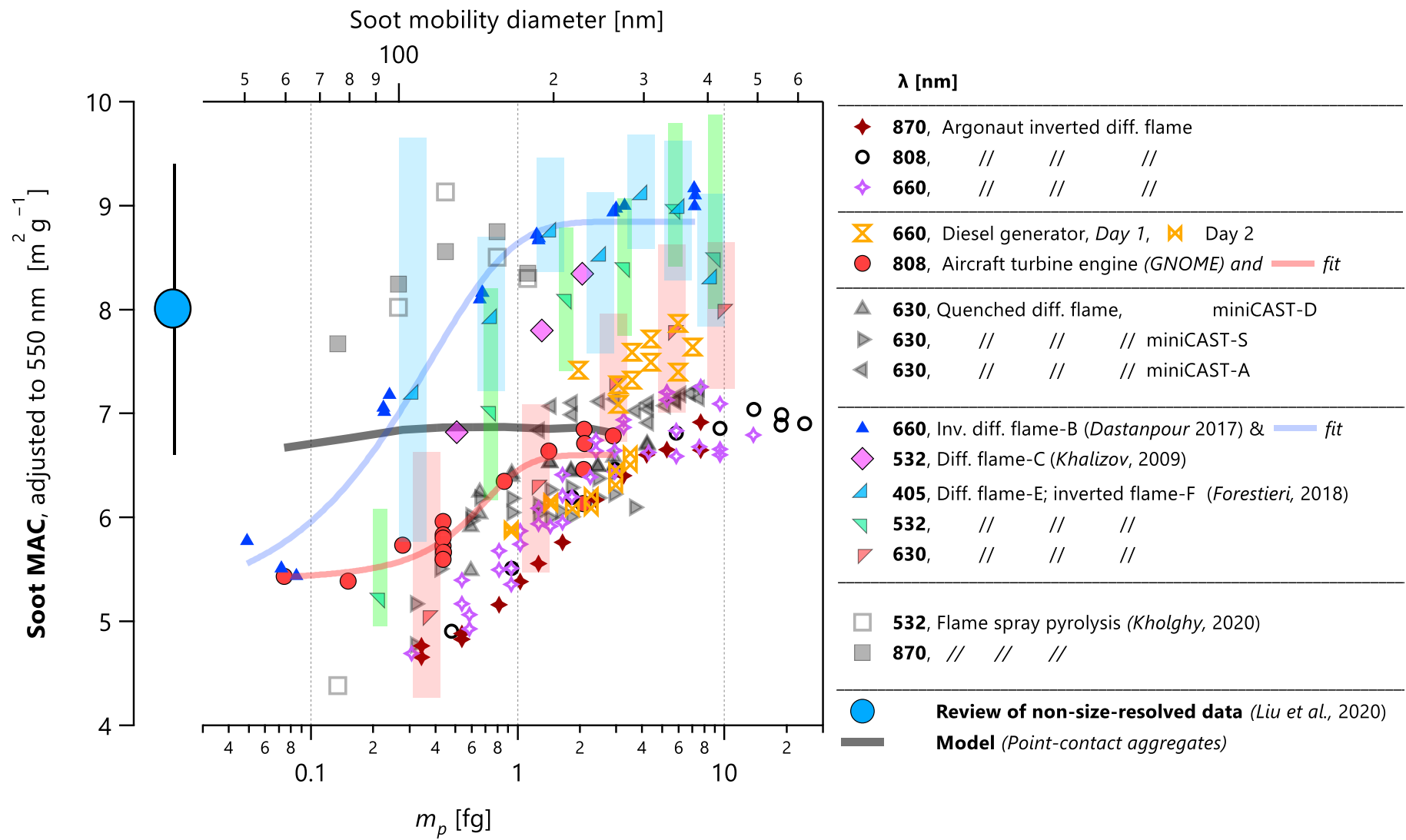

Figure S7. Similar to Figure 5 but with data overlaid on a single panel.

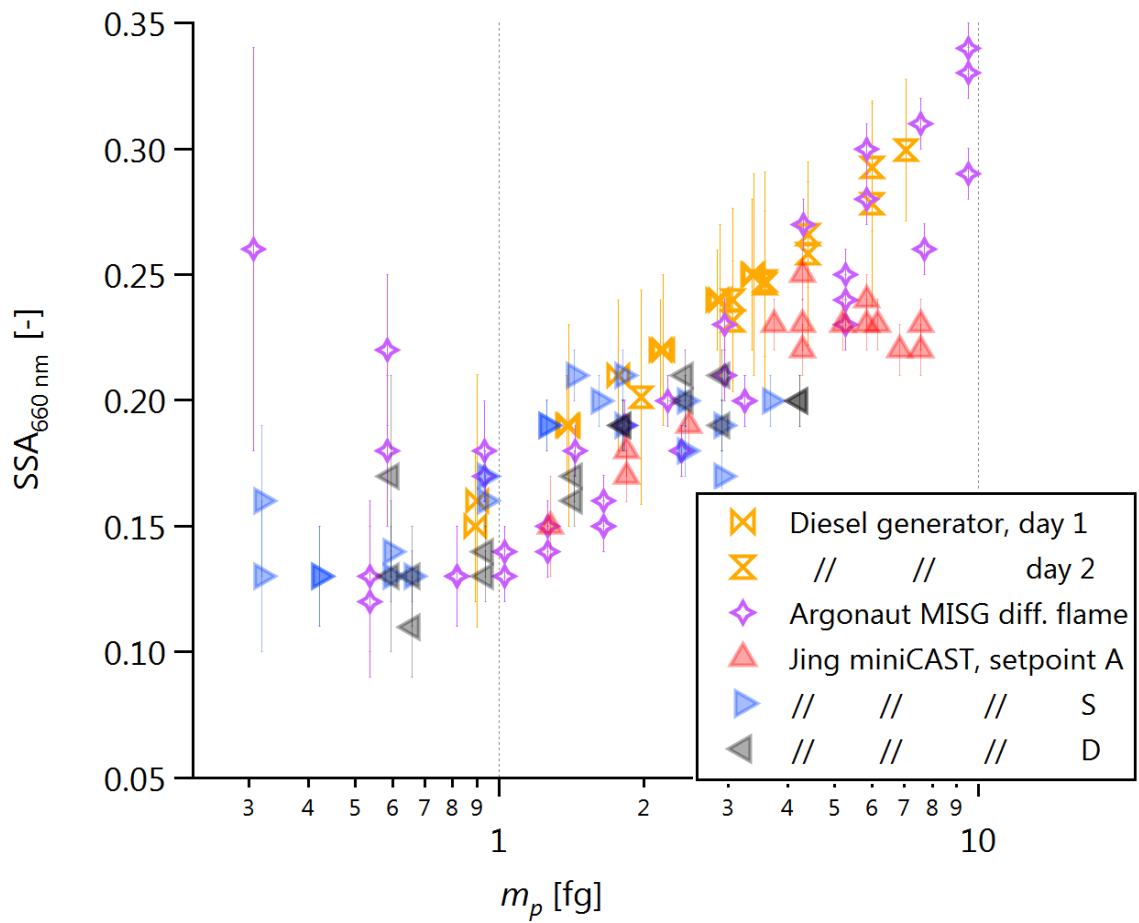

Figure S8. Single scattering albedo (SSA) versus single-particle mass for the 660 nm CAPS PM$_{SSA}$ data.

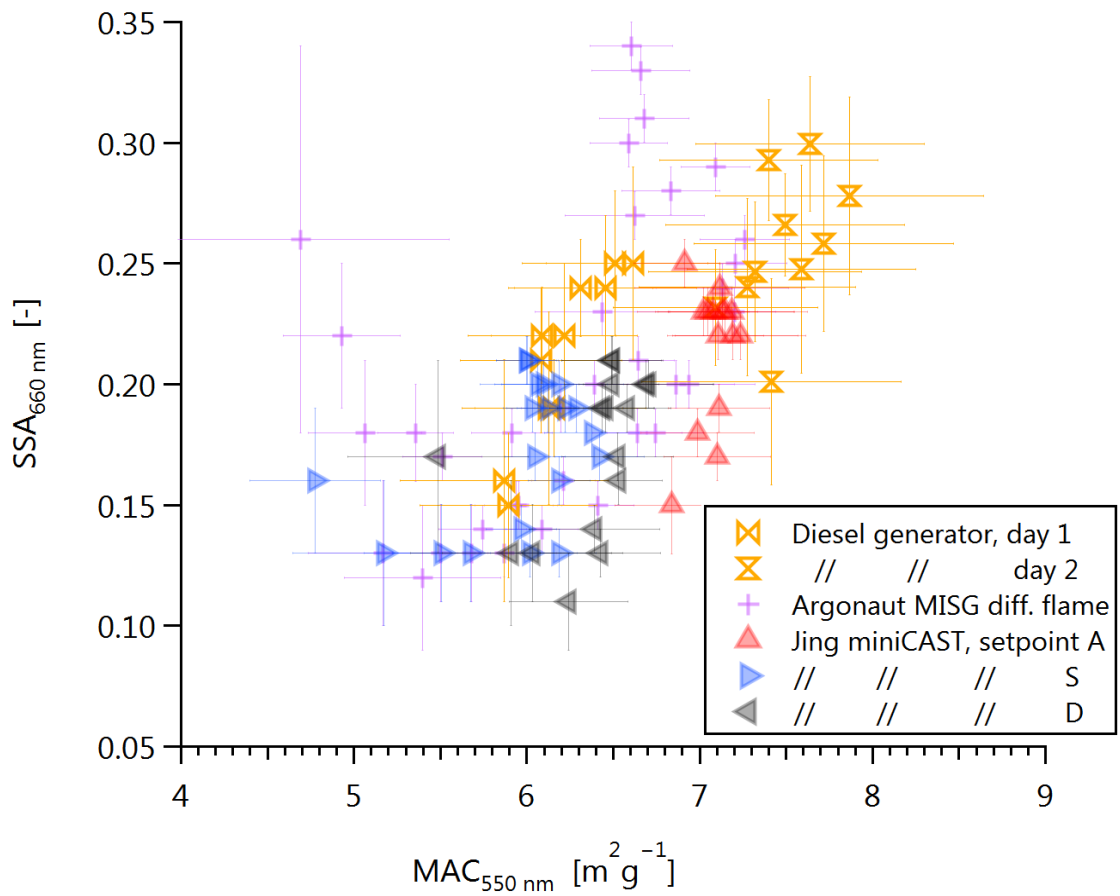

Figure S9. Single scattering albedo (SSA) versus $MAC_{550nm}$ for the 660 nm CAPS $PM_{SSA}$ data.

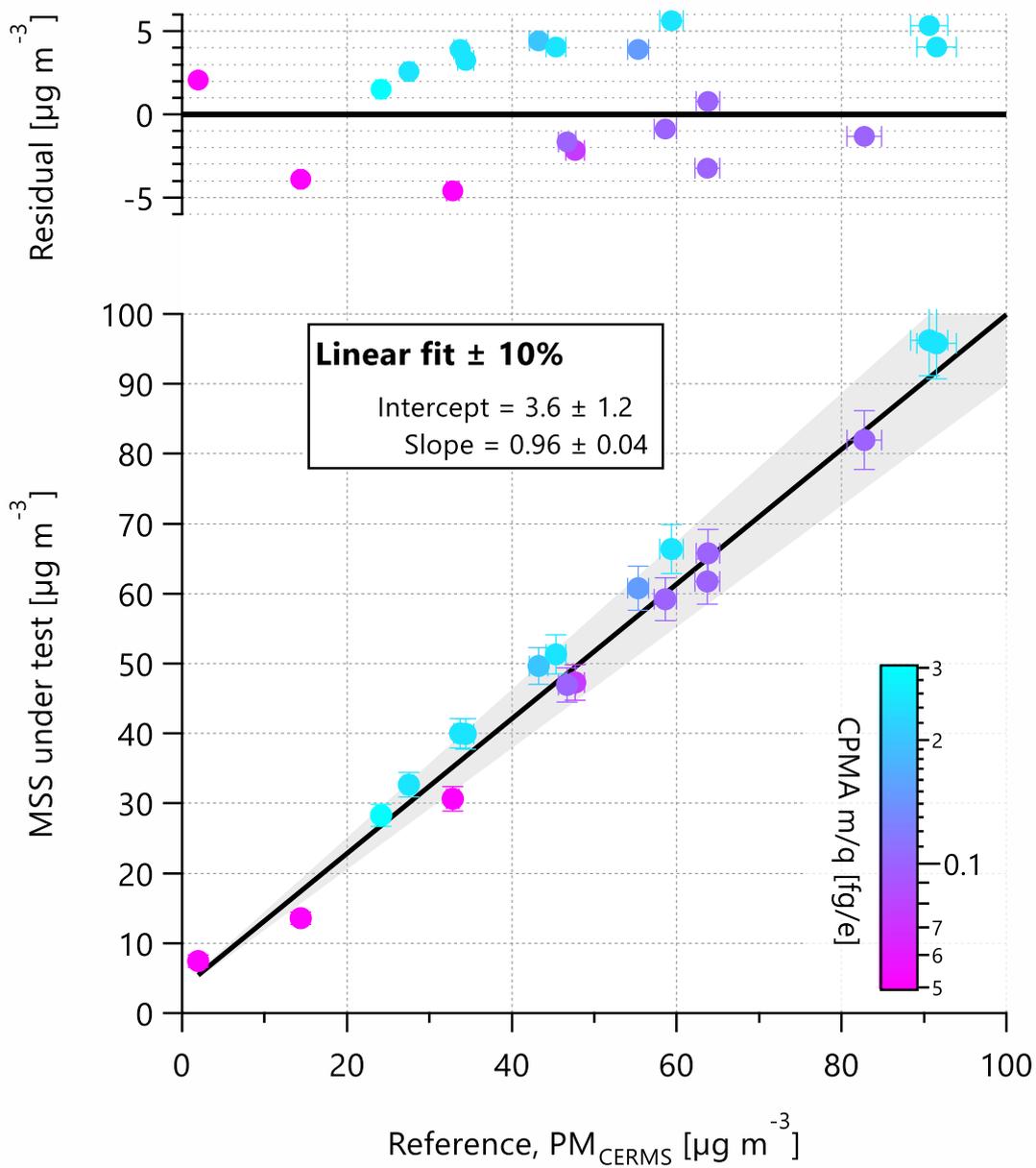

Figure S10. Calibration of the MSS used in this study using the Gnome data shown in Figure 3, Figure S6, and other figures. The size-resolved MAC signal results in a scatter of only approximately 10% in the calibration scatterplot and residuals. However, this scatter is systematic enough to allow for the size-resolved MAC to be inferred, as discussed in the manuscript.